\RequirePackage{fix-cm}
\documentclass[smallextended]{svjour3}       
\smartqed  
\usepackage{graphicx}
\usepackage{algorithm}
\usepackage{algpseudocode}
\usepackage{amsmath}
\usepackage{amssymb}
\usepackage{multirow}
\usepackage[title]{appendix}
\usepackage[usenames]{color}
\usepackage{float}

\usepackage{ulem}

\algnewcommand\algorithmicforeach{\textbf{for each}}
\algdef{S}[FOR]{ForEach}[1]{\algorithmicforeach\ #1\ \algorithmicdo}

\newcommand{\subsubsubsection}[1]{\paragraph{#1}\mbox{}\\}
\begin{document}

\title{A Decentralised Self-Healing Approach for Network Topology Maintenance}


\author{Arles Rodr\'iguez \and 
        Jonatan G\'omez \and 
        Ada Diaconescu
}


\institute{A. Rodr\'iguez \at
              Fundaci\'on Universitaria Konrad Lorenz\\
              Tel.: +57 3133731086\\
              \email{arlese.rodriguezp@konradlorenz.edu.co} 
           \and
           J. G\'omez \at
              ALIFE Research Group - Universidad Nacional de Colombia \\
              \email{jgomezpe@unal.edu.co}
           \and 
           A. Diaconescu \at
              Telecom Paris \\
              \email{ada.diaconescu@telecom-paris.fr}
}

\date{Received: date / Accepted: date}

\maketitle

\begin{abstract}

In many distributed systems, from cloud to sensor networks, different configurations impact system performance, while strongly depending  on the network topology. Hence, topological changes may entail costly reconfiguration and optimisation processes. 
This paper proposes a multi-agent solution for recovering networks from node failures. {To preserve the network topology, the proposed approach relies on local information about the network's structure}, which is collected and disseminated at runtime. 
The paper studies two strategies for distributing topological data: {one based on} Mobile Agents (our proposal) and {the other based on} Trickle ({a} reference gossiping protocol from the literature).  {These two strategies were adapted for our self-healing approach  -- to collect topological information for recovering the network; and were evaluated in terms of resource overheads.} 
Experimental results show that both variants can recover the network topology, up to a certain node failure rate, which depends on the network topology. At the same time, Mobile Agents collect less information, focusing on local dissemination, which suffices for network recovery. This entails less bandwidth overheads than when Trickle is used. Still, Mobile Agents utilise more memory and exchange more messages, during data-collection, than Trickle does. These results 
validate the viability of the proposed self-healing solution, offering two variant implementations with diverse performance characteristics, which may suit different application domains.


\end{abstract}

\section{Introduction}
\label{intro}

Large-scale distributed systems, including electric power grids, the Internet, social networks, data centres and clouds, feature different network topologies -- e.g. random, scale-free, small world or community  \cite{Hayashi2016}. Hence, several researches have identified and studied different types of network topologies \cite{Watts1998,White2005,Small2016,Takemoto2012}, aiming to determine their impact on various system functions and properties, including the speed and robustness of data collection \cite{Rodriguez2016,Chen2018,levis2011trickle}, information replication \cite{Weng2015}, or generation of resilient networks via self-organisation techniques \cite{Hayashi2016,Debbabi2012}.

In many distributed applications, such as server farms and cloud management, maintaining and recovering servers from failures is essential for meeting stringent requirements of availability and reliability. Since these systems serve millions of users in parallel, downtimes can generate losses of thousands of dollars per minute \cite{Lalanda2013}. System administration expertise is expensive and in short supply, considering the pervasiveness of a distributed system. Moreover, experts may be unable to react fast enough for dealing with multiple failures within short periods, e.g. minutes \cite{Lalanda2013}. E.g., in 2017, a disruption in the Amazon s3 service occurred due to a command entered incorrectly that removed a large set of servers causing failures in other important subsystems \cite{amazon2017}. Hence, fast automated solutions are needed to enable system \textit{self-healing}.

Moreover, many distributed graph algorithms -- e.g. spanning tree calculation \cite{Rodrigues2018}, vertex colouring  \cite{Kuhn2020} or leader election \cite{ghaffari2013near} -- depend on the underlying network topology and incur important overloads to re-execute if the topology changes. 
Network topology is also essential for information spreading, hence impacting traffic dynamics and performance \cite{VanDerHofstad2016}. Furthermore, topologies can be optimised for maximising network transmission capacity \cite{Chen2018}.  
Relevant application domains may range from clouds and data farms, to sensor networks and the IoT. In such contexts, self-healing solutions must  maintain the network's topology as similar as possible to the original one. 

This paper proposes a {generic} network self-healing approach based on two decentralised processes: i) a data-collection and dissemination algorithm for gathering knowledge about the network topology; and, ii) a local node recreation and re-connection mechanism for recovering the network topology when nodes fail.

Concerning data-collection, we adopt a Mobile Agent based algorithm proposed in previous work \cite{Rodriguez2019}, and adapt it to the self-healing problem. {Here, Mobile Agents move from node to node and disseminate the network topology data}. We evaluate {the} performance {of this approach} with respect to a well-known algorithm from the literature -- Trickle \cite{levis2011trickle} -- for data propagation and maintenance in sensor networks. 

In the proposed {self-healing} approach, {network} {nodes are} {modelled as} {static agents that learn about neighbouring nodes dynamically}, {via} {either Trickle or the Mobile Agents process.} Nodes also check and detect each other's failures. {A failure corresponds to a} {node} {crash,} {implying that the node stops its processing and is disconnected from the network. Failures occur with a certain {probability} rate{, as} defined in subsection \ref{subsec:entity-types}. In each {simulation} round,} 
{each node identifies the differences between its local knowledge about the network topology and the perception it has of its neighbouring nodes (of degree one), which it checks upon via a heart-beat message. When a failed node is detected by its neighbouring nodes, these latter coordinate locally to recreate and reconnect the missing node, based on their local topology knowledge.  Concrete simulation time parameters are defined in subsection \ref{subsec:scheduling}.}




Via a wide range of simulations, the paper analyses and compares the two algorithms, adapted for network self-healing, in terms of resource consumption (i.e. communication bandwidth, message numbers and memory overheads) and ability to recover the network from node failures. 
{Results show that the network can recover its topology, up to a maximum node failure rate, provided that the remaining nodes were able to acquire information about the entire network topology beforehand. To ensure this in the initial experiments, we started removing nodes {only} after an initial  {data-}acquisition period, estimated to suffice for collecting {knowledge of the entire} network topology. In further experiments we showed that the network could also recover {completely even} if node failures started before {knowledge of} the entire network topology could be acquired {locally, by each node}. This is possible because}{, by their nature, both data-collection processes acquire \textit{local} topological knowledge first, then extend this knowledge progressively to the entire network. As the self-healing process uses local topological information, the extended knowledge about the entire network is superfluous.} {As previous work highlighted the significant impact of network topology on the performance of decentralised algorithms (e.g. data-collection \cite{Rodriguez2019}), experiments were carried-out on a diverse selection of network topologies (i.e. shown to make a difference in \cite{Rodriguez2019}).}


To stress-test the node recovery algorithm, separately from the data collection process, we first performed experiments where all nodes received information about the entire network topology at initialisation time. This allowed to establish the maximum node failure rate at which the proposed self-healing  approach was still viable (i.e. able to recover the entire network). 
Further experiments established the efficiency of the two data-collection algorithms with respect to this self-healing process. The evaluation comparison between the proposed Mobile Agent technique and the base-line approach provided by Trickle shows that Mobile Agents are comparable to Trickle in terms of topology maintenance, while incurring lower bandwidth overheads. {Still, when mobile agents are used, they incur slightly more memory overheads. This is because agents represent extra entities that move across the network, carrying topology information and exchanging extra messages with the nodes.} Highlighting such trade-offs allows system designers to adopt the most suitable data-collection method for their particular application context. 

%

The paper's contributions include:
\begin{itemize}
\item Proposing a {multi} agent-based approach for self-healing network topologies, {based on local topological data-collection; 
\item Adopting two decentralised data-collection and dissemination algorithms: Mobile Agents (based on previous work \cite{Rodriguez2019}) and Trickle (reference gossip protocol from the literature), and adapting them to achieve network self-healing;}
\item Evaluating the Mobile Agents algorithm and comparing it to the baseline approach (Trickle). This analysis highlights the two algorithms' self-healing abilities and resource overheads, for various node failure rates and different network topologies (i.e. diverse degree distributions). It highlights their strengths and weaknesses pointing to their applicability within different contexts; 
\item Showing that the proposed Mobile Agents  approach, compared to the Trickle baseline, offers similar self-healing performance while incurring lower communication overheads. While Mobile Agents consume more memory than Trickle while gathering information, they store less neighbourhood information on network nodes and hence use less memory once the collection process terminates.

\end{itemize}

The insights obtained via the presented evaluation indicate that, in the targeted application context, network self-healing is possible without collecting information on the whole network topology. In other words, some of the topological information collected (e.g. by Trickle) is not employed in the self-healing process, hence incurring unnecessary resource overheads. The Mobile Agents approach somewhat circumvents this issue by focusing on local information collection, limited to network neighbourhoods, which is essential for topology repairs. Experimental results establish the maximum node failure rate for which the network topology can still be recovered, via the two algorithms.  Future work will concentrate on determining the (minimum) extent of neighbouring network information that is necessary for ensuring network recovery, for different network topologies and failure rates.

To describe our contribution, we employ the ODD (Overview, Design Concept and Details) Protocol \cite{Grimm2010}, and associated documentation suggestions \cite{Grimm2013}. The ODD protocol facilitates the understanding and clarifies the documentation of agent-based simulation models (ABMs). It starts by defining the purpose of the model, then highlights the main theoretical concepts and their transfer to simulation design, and finally  provides the implementation and simulation details.

The remainder of this paper is organised as follows. Section \ref{sec:relatedwork} presents related work and highlights the novelty of our proposal. Section \ref{sec:odd} describes the proposed self-healing solution and its two data-collection variants, following the ODD protocol. Section \ref{sec:experiments-and-results} presents experimental results and discusses their significance.  
Finally, section \ref{sec:conclusions} concludes the paper and points to future research directions.

\section{Related Work}
\label{sec:relatedwork}

The proposed approach mainly differs from existing work in that it aims to recreate and maintain a network's topology, rather than merely its connectivity or diameter. This is feasible in application contexts where nodes can be recreated and reconnected (e.g. spare servers in Data Centres; dormant devices in sensor networks; redundant robots or handhelds in ad-hoc mobile networks).   
For instance, \cite{Hayashi2016} proposed a design method for growing both robust and efficient onion-like topological structures. \cite{Bai2019} aimed to build a reliable topology based on a fractal cell-structure and compare its designed network with scale-free networks defined following the model of \cite{Barabasi2003}. This is complementary to our approach in that we aim to recover the network topology in case of failure. Other approaches aimed to deal with node failures by reconnecting remaining nodes; hence avoiding network splitting into isolated components, which would render the system dysfunctional. The proposals in \cite{Gallos2015} and \cite{WANG2017} for instance trigger the node reconnection process to maintain connectivity when a node loses a certain number of neighbours; or, to maintain inter-node distances below a determined maximum distance, or as short as possible compared with the original network. As before, this approach does not recover failed nodes; and hence does not preserve network topology. 

Another related approach \cite{Safaei2020} studies complex network robustness and proposes a rewiring mechanism based on Shannon entropy to reconnect a network and increase its resilience. 
The declared objective of this approach is to protect the existing networks against the possibility of targeted attacks or random failures. This is also complementary with our approach, which can maintain a targeted topology once this was established to ensure resilience. 

\cite{Chaoqi2018} brings to the fore the importance of effective node repair strategies for complex networks. It uses an energy transfer function to describe the cause of cascade failures triggered by a failed node. 
According to this model, a failed node causes its load to be delegated onto adjacent nodes.  
However, if this extra load is greater than the initial load of a node, which is correlated with its node degree \cite{Wang2008}, then a cascade failure may occur. The proposal provides a node repair sequence and indicates that this should be perfoby maintenance staff. In our case, the node recreation algorithm can be updated to follow this sequence.    

\cite{Ochoa-aday} proposed a self-healing protocol for Software Defined Networks (SDNs). This work aimed to maintain a given topology in two steps: first, using multi-cast for network discovery and state data collection; and second, using an autonomic failure recovery mechanism. Experiments were performed for Scale-free networks. Our approach goes deeper into exploring and evaluating alternative data-collection mechanisms (based on Trickle and Mobile Agents), to reduce resource overheads. 
{We adapt Trickle and Mobile Agents algorithms to provide nodes with topological data, as necessary to maintain a targeted network topology. Moreover, we analyse the efficiency of these algorithms for different topologies, showing that structural characteristics (though often ignored) do matter.}

{Trickle is a scaleable and robust algorithm for propagating and maintaining information in low-power, lossy networks (e.g. wireless sensor networks). It was defined under the RFC6206 standard \cite{levis2011trickle}, with common applications including traffic timing control, multi-cast propagation and route discovery. 
It uses a ``polite gossip'' algorithm, where nodes broadcast local information to neighbours periodically, but stay quiet if they have seen that same information pass-through recently. To avoid flooding the network, each node adjusts its re-transmission rate depending on the perceived level of information novelty. We adopt Trickle as a baseline for evaluating our Mobile Agents approach.}

{In short, Trickle executes on each local node, as follows \cite{levis2011trickle}. First, a time \textit{interval} is defined (in milliseconds). For each interval, a \textit{redundancy counter} is initialized to  zero. Afterwards, whenever Trickle hears a transmission that is consistent (i.e. redundant with respect to previous transmissions), it increments the counter. At a predetermined time in the interval, Trickle transmits if and only if the redundancy counter is lower than a predefined redundancy constant. When the time interval expires without transmission, Trickle doubles the interval. In case of an inconsistent transmission (i.e. new data with respect to previous transmissions), Trickle reduces the interval time to a predefined minimum value.} 

{In this paper, we adapted Trickle to work with discrete simulation rounds instead of milliseconds; specified the transmitted information (i.e. the topology data); and defined a consistent transmission (i.e. receiving topology data that is already available locally) versus an inconsistent transmission (i.e. receiving new topology data).}

{\cite{Rodriguez2016} and \cite{Rodriguez2019} studied data collection in Complex Networks with failure-prone Mobile Agents. In these works, failures intervened at the mobile agent level (i.e. lost agents). 
Experimental results in \cite{Rodriguez2016} showed that network topology impacts data-collection speed. \cite{Rodriguez2019} proposed an agent movement algorithm (called `random with marks') that performed well in a variety of complex network topologies. This algorithm added a `visited' flag to nodes after a mobile agent passed through them. It also allowed Mobile Agents located on one node to obtain the `visited' status of neighbouring nodes and hence to avoid revisiting previously explored nodes.}

{In the present paper, Mobile Agents move through the network based on the `random with marks' algorithm to collect and distribute topological information
(Cf. sec \ref{subsubsubsec:mobile-agents-submodel}). 
We additionally imposed that if a node fails then all the Mobile Agents located on this node fail as well. To reduce network consumption, if a Mobile Agent does not collect any new information after visiting a predefined number of nodes, then it stops its execution to release memory and reduce bandwidth overheads.
}

\section{Network Self-healing Description, based on ODD Protocol}
\label{sec:odd}

\subsection{Purpose}
\label{sec:overview}

The purpose of the proposed model is to evaluate our self-healing approach for maintaining a network's topology as close as possible to the original one, in case of node failures. The model offers two implementation variants differing in their data-collection and dissemination method: one adopted from our previous work (Mobile Agents) and one from the literature (Trickle) -- both adapted to the targeted self-healing application. Experimental simulations aim to compare the two variants and highlight their advantages and disadvantages in terms of network self-healing abilities and resource overheads. The ultimate goal is to show that topological self-healing is possible via decentralised algorithms, provided that topological knowledge can be gathered in time.

\subsection{Entities, State, Variables and Scales}
\label{subsec:entity-types}

The simulation model defines three types of entities: one external entity representing the \textit{network environment} and two agent types -- \textit{static nodes} and \textit{mobile agents}.

The \textit{network environment} represents the network's state during runtime. It defines the network's topology as a graph, which allows evaluating its similarity with the original network topology, to be maintained. The network environment representation also indicates whether a node is \textit{alive} or in a \textit{failed} state. The environment is parametrised with a \textit{failure probability} $p_f$, which determines the likelihood of node failures at each round. With respect to data-collection, the network environment is configured to employ one of the available techniques: Trickle or Mobile Agents. Finally, the network environment maintains a \textit{round number} that represents the agents' discrete execution step. Each step corresponds to a state-change in a real system, considering that sufficiently fine-grained simulation steps would approximate a continuous real-time process. 
  

\textit{Nodes} are static agents that represent the different processes (or platforms) of a selected network. Each node has: a unique \textit{id}; a \textit{visited} status flag; a \textit{message queue} for communicating with other node agents; \textit{memory} space for storing topology data; and an \textit{algorithm} (or program) for processing incoming messages, simulating the node's failure, detecting missing neighbours and recreating them. This algorithm program is executed at every round, or simulation step. We implemented three variants of this node program, varying in the way in which network topology information is acquired. In the first variant, all topology information is provided statically; in the second one, topology information is propagated via the Trickle algorithm; and, in the third one, via the Mobile Agents approach. Unique node id-s can be ordered sequentially -- this property is employed to ensure the uniqueness of node recreation when failed nodes have several neighbours (Cf. \ref{subsubsec:self-healing}). More sophisticated leader election approaches may be adopted instead (future work). 

In the Mobile Agents variant, each \textit{mobile agent} is endowed with: \textit{memory space} to store data collected from nodes on the network topology; and an \textit{algorithm} (or program) to communicate with visited nodes, simulate agent failures, and select the next node to visit. The agent program is executed at every round, or simulation step.

In both nodes and mobile agents, topology data is represented as a dictionary $(key: value)$, where the \textit{key} is the node $id$, and the corresponding value is a list of neighbours of this node. E.g., for a network a---b---c, a node or mobile agent with all the topology data stores this information as: $\{(a:\{b\}), (b:\{a, c\}), (c:\{b\})\}$.

\subsection{Process overview and scheduling}

\subsubsection{Scheduling and synchronisation}
\label{subsec:scheduling}
Simulation time is defined via discrete rounds, or steps. In each round, the network environment updates statistics stored about the consumed resources: memory (in bytes) used by nodes and by agents, communication bandwidth (in bytes), and number of messages received by nodes. It also determines which nodes fail, with a probability $p_f$, corresponding to node crashes in real systems. 
Agents are implemented in a multi-agent system simulator (multi-threaded)\footnote{Source code is available at: https://github.com/arleserp/NetworkRecoverySim} running on a single machine. In each round, each agent executes its program once. Nodes and mobile agents run in parallel, in random order, and are synchronised at the end of each step, for collecting metrics.

\subsubsection{Self-healing process overview}

The proposed network self-healing approach relies on two decentralised mechanisms. Firstly, nodes collect and distribute information about the network topology -- e.g. via the proposed Mobile Agents algorithm, or via the adopted Trickle standard. Secondly, nodes detect their neighbours' failures, and consequently recreate and reconnect them using local topology information and local interactions.

\subsubsection{Data-collection and dissemination}

Data collection and dissemination depends on the selected algorithm. 

\textit{Trickle} runs in each node and uses a ``polite gossiping'' approach to disseminate topological information. That is, each node receives information from (some of) its neighbours, at a certain time interval. It re-broadcasts the information \textit{iff} it is sufficiently novel with respect to information seen recently. To establish information novelty, Trickle uses a \textit{redundancy counter} for keeping the number of times, in a row, that identical information has been seen. This counter is reinitialised to zero every time novel information is received. Once the counter reaches a maximum threshold $k$, redundant information is discarded. The redundancy threshold and the various time intervals for rebroadcasting information are essential configuration parameters for the Trickle approach.  

\textit{Mobile Agents} explore the network using a movement algorithm (developed in previous work \cite{Rodriguez2019}, Cf. subsec. \ref{subsubsubsec:mobile-agents-submodel}). The algorithm promotes random movement, while also marking and avoiding previously visited nodes. This was shown to be a suitable strategy for a wide variety of network topologies; whereas other strategies were slightly better for some topologies yet worse for others (where the differentiating topological factor was the standard deviation of the network's {betweenness} centrality distribution). 
As agents move across the network, they collect topological data from visited nodes; and deposit collected data onto visited nodes. Data collection and depositing is based on the union between the visited node's data and the agent's carried data. 

\subsubsection{Node recovery process}

To maintain the network topology, failed nodes are replaced by node replicas, which are then connected to the remaining nodes and initialised with topological information available locally. 
{In our experimental simulation, nodes are modelled as static software agents and so we simply recreate new agent instances to replace failed ones (i.e. failed nodes). In concrete applications, the creation of node replicas will depend on the node types and domain-specific constraints, case-by-case. For instance, in a software system based on a micro-service oriented platform 
\cite{Aderaldo2017}, it is possible to create new software instances when failures occur in micro-services or in the containers managing them. 
}
{Similarly, in data centres or cloud computing platforms, physical nodes such as failed servers can be replaced by switching-on new ones, kept in stand-by. In more dynamic scenarios, like mobile sensor networks or robotic systems, new nodes might have to be brought on site, e.g. \cite{10.5555/3398761.3398809}. 
From a generic perspective, the main aspect differentiating these scenarios is the cost and delay induced by node replication. In all cases, for the self-healing to be viable, the self-repair process must be faster than the node failure rate}.     

Self-healing actions are triggered and enacted by neighbouring nodes, based on local topology information, collected previously. Upon creation, node replicas are also initialised with local topological information from neighbouring nodes. When Mobile Agents are used, a new agent is also created for each new node, to compensate for agents that were potentially lost during the node's failure {(i.e. if a node fails, Mobile Agents located on this node also fail).} 

Each node is provided with an algorithm for recovering the network structure. It assumes that each node can sense its immediate neighbours. At each round $t$, a node's perception of its neighbours is compared to memorised data about the local topology, stored at $t-1$. Differences are interpreted as node failure(s) and the missing node(s) are replaced by new one(s). 

Consider a network with 2 interconnected nodes, $a$--------$b$. If node $a$ fails it can only be recreated if node $b$ has knowledge of its previous existence as a direct neighbour. This suffices for recovering the network from a single node failure. Nonetheless, simultaneous failures of interconnected nodes require nodes to monitor neighbours at multiple hops. For instance, considering a network of 3 nodes, $a$--------$b$--------$c$, where each node only monitors immediate neighbours. If nodes $a$ and $b$ fail simultaneously (within the same simulation round), then $c$ can recreate $b$ but not $a$. To tackle this problem, $c$ must also collect information about $a$-- hence {information} about two-hop neighbours {are required in this example to recover this network.} Then $c$ can recreate $b$ and send its information about its direct neighbour $a$. In this way, $b$ can recreate $a$ and the entire network can be recreated based on local information and interaction. 
In the same example, $a$--------$b$--------$c$, if $b$ fails then either $a$ or $c$ can recreate it. To avoid duplicate replicas, the neighbour with the minimum node $id$ recreates a missing node -- hence, $a$ recreates $b$.

\subsection{Design concepts}

An important aspect that experiments take into account is that the performance of the data-collection algorithm depends on the network topology (more precisely on its betweeness centrality metrics \cite{Rodriguez2019}). Therefore, to run the experiments, we selected different topologies, with varying degree and centrality distributions (identified to make a difference in previous work). An important assumption is that the number of interconnected nodes that fail simultaneously is limited, so that the network can recover faster than the nodes fail -- e.g. as in \cite{Majdandzic2016}. Similarly to living organisms, a network can self-repair after relatively small distributed failures, but not after multiple major failures happening at once. When a node fails, it is deleted from the network along with its connections and visiting agents.

Another important assumption is that a node can recreate and/or reconnect to any other node, as needed to rebuild the network topology. It is also assumed that nodes fail independently (with a probability $p_f$ in the simulation), which generally holds in computer networks. Nonetheless, this assumption would have to be reconsidered in other application domains, such as financial markets or organisms, where systems contain multiple interacting networks, making node failures inter-related. For instance, \cite{Majdandzic2016} defines three types of failures:\textit{ internal failures}, associated with organ malfunction or mismanagement; \textit{external failures}, resulting via propagation from other failed nodes, within the same network; and, \textit{dependent failures}, produced by cascading failures from another connected network. Our work aims to self-heal internal node failures.


To be applicable to large-scale distributed systems, the proposed solution must scale with the number of network nodes, and, to a certain extent, with the frequency of topological changes. In terms of adaptation, redundancy of the collected data is a key factor to reduce resource consumption. Mobile agents mark their environment (i.e. visited nodes) to improve network exploration. They are dropped (i.e. removed) if the information they carry is identical to the one held by a visited node. Similarly, the Trickle variant drops messages if their information is the same with the one detained by the receiving node.

\subsection{Details}


\subsubsection{Initialisation}

The simulation starts by initialising its three main entity types (Cf. subsec. \ref{subsec:entity-types}): network environment, static nodes and mobile agents (when used).  

\subsubsubsection{Network Environment}

\begin{figure}[H]
\centering
\includegraphics[scale=0.4]{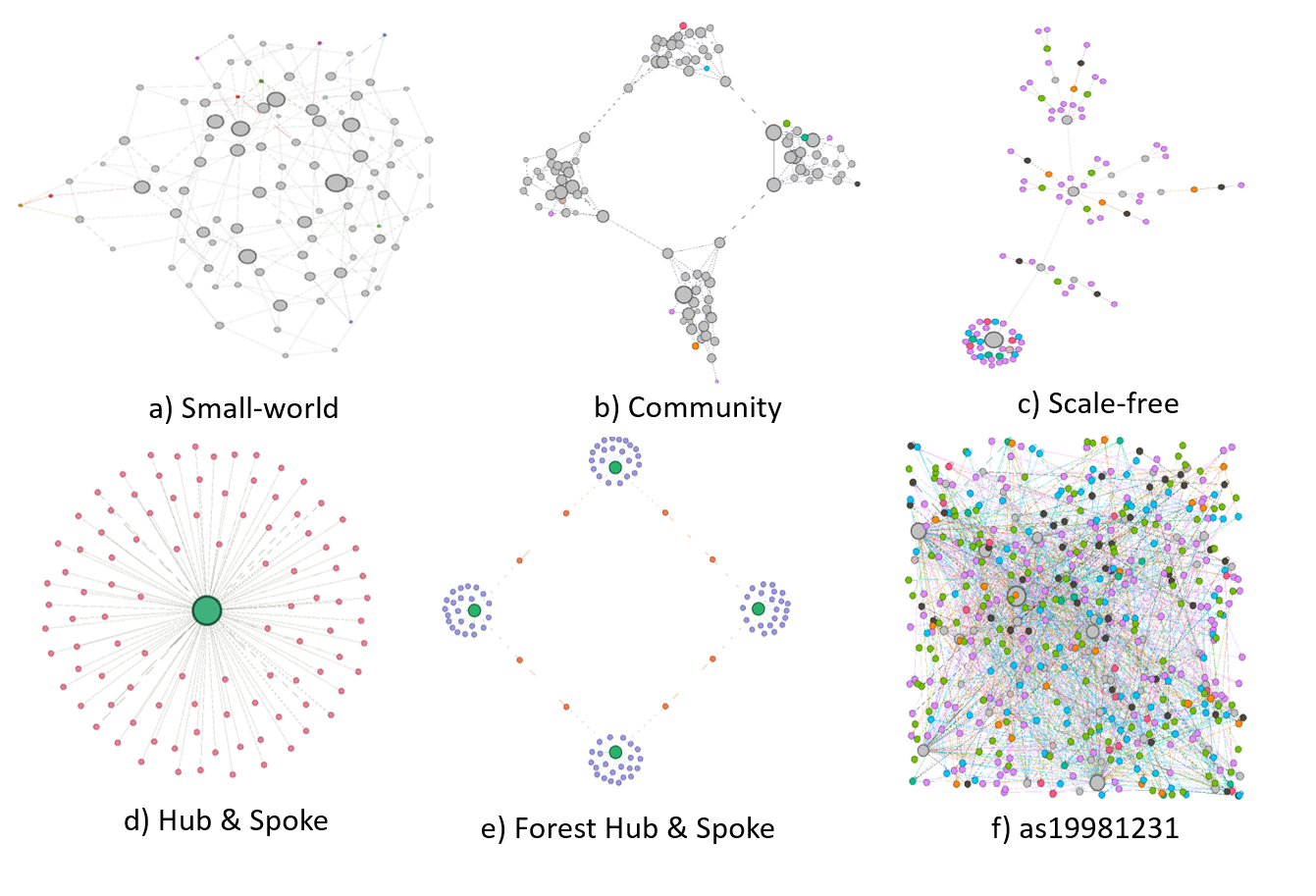}
\caption{Network topologies selected for experiments -- drawn with Gephi tool \cite{ICWSM09154}}
\label{fig:all-networks}       
\end{figure}

The initialisation of the network environment requires selecting 
the network topology type (from a predefined set) and the 
simulation time (in number of rounds). 
Several network topologies were defined for testing the proposed approach, as shown in Figure \ref{fig:all-networks}. These topologies were labelled according to their general degree distributions \cite{VanDerHofstad2016}: 
  \textit{Small-world}, \textit{Community}, \textit{Scale-free}, \textit{Hub \& Spoke} and\textit{ Forest Hub \& Spoke}. The selection was based on the results in \cite{Rodriguez2016}, where these networks were shown to be the most challenging for data collection when using our Mobile Agents approach -- this was due to the increased network {betweenness} centrality of certain nodes, or \textit{hubs}, relative to the other nodes. For comparability reasons, these networks consisted of about 100 nodes.  Additionally, to test our proposal on a larger-scale realistic network, we selected from the SNAP database a network representing Internet routers sub-graphs (i.e. network \textit{as19981231}, with 512 nodes) \cite{leskovec2005graphs}. 
  
\textit{Small-world} networks are characterised by the fact that the distance between any two randomly selected nodes is relatively small (i.e. proportional to the $log$ of the total number of nodes). A good example of small-world network is the Internet because typical distances between any two points are relatively small~\cite{VanDerHofstad2016}. 
The Small-world network in Fig. \ref{fig:all-networks}-a was generated via the Watts-Strogatz model \cite{Watts1998}. This model starts with a lattice network of $n$ vertices with degree $k$ and subsequently rewires the nodes with a probability $\beta$. 
A  probability $\beta$ is defined to rewire a lattice network of $n$ vertices with degree $k$.  Hence, $\beta = 0$ generates a regular network, and $\beta = 1$ a random network; in-between values $\beta=(0,1)$ generate Small-World networks \cite{Mori2015}.  

\textit{Community} networks correspond to structures where nodes can be assigned to different clusters. Each cluster is highly interconnected internally and has relatively few links among nodes belonging to different clusters \cite{Girvan2002}. Examples  include certain social networks, collaboration networks and web pages with related topics~\cite{Girvan2002}. 
Fig \ref{fig:all-networks}-b depicts a Community network generated by defining four clusters, each one representing a small-world with its own $k$, $\beta$, and $n =m/n_{clusters}$, where $m$ is the number of nodes in the network. The four clusters are interconnected via a circle, consisting of node pairs selected randomly from each cluster. 

\textit{Scale-free} networks are characterised by degree distributions that follow a power law \cite{Barabasi2003}. That is, a few nodes, or hubs, have a number of connections that greatly exceeds the average connections of the other nodes. This makes these networks highly robust to accidental failures but susceptible to targeted node attacks. Examples include social networks, financial networks, some computer networks and protein interaction networks. Scale-free networks (e.g. Fig. \ref{fig:all-networks}-c) are typically obtained via a \textit{preferential-attachment} process. Namely, starting with $sn$ nodes and $\eta$ connections, each new node is further connected to the existing nodes via $\eta$ links with a probability $p_i=\frac{k_i}{\sum_j(k_j)}$ which is proportional to the degree $k_i$ of each existing node $i$ \cite{White2005,Small2016,Takemoto2012}.  

\textit{Hub \& Spoke} networks feature a star configuration of at least one cluster, each one containing $n$ nodes -- one central node and $n-1$ adjacent nodes (e.g. Fig. \ref{fig:all-networks}-d). We selected this topology to `exaggerate' the node degree differences, with the hub being very well connected and all other nodes being weakly connected (one or two links). This is because our tests showed that weakly connected nodes were easier to loose (i.e. no recovery, Cf. Fig. \ref{fig:missingnodes} and Table \ref{tbl:success-allinfo}). This topology can occur for instance in cloud systems, where central servers require extensions via further instances for handling increasing workloads 
\cite{Mahmood2011}. A \textit{Forest Hub \& Spoke} network was defined in this paper via the union of 4 clusters each one with a Hub \& Spoke network (e.g. Fig. \ref{fig:all-networks}-e).

The larger-scale routing network (512 nodes), identified as \textit{as19981231}, or `AS' in short, is depicted in Fig. \ref{fig:all-networks}-f.  

\begin{figure}[H]
\centering
\includegraphics[width=\textwidth]{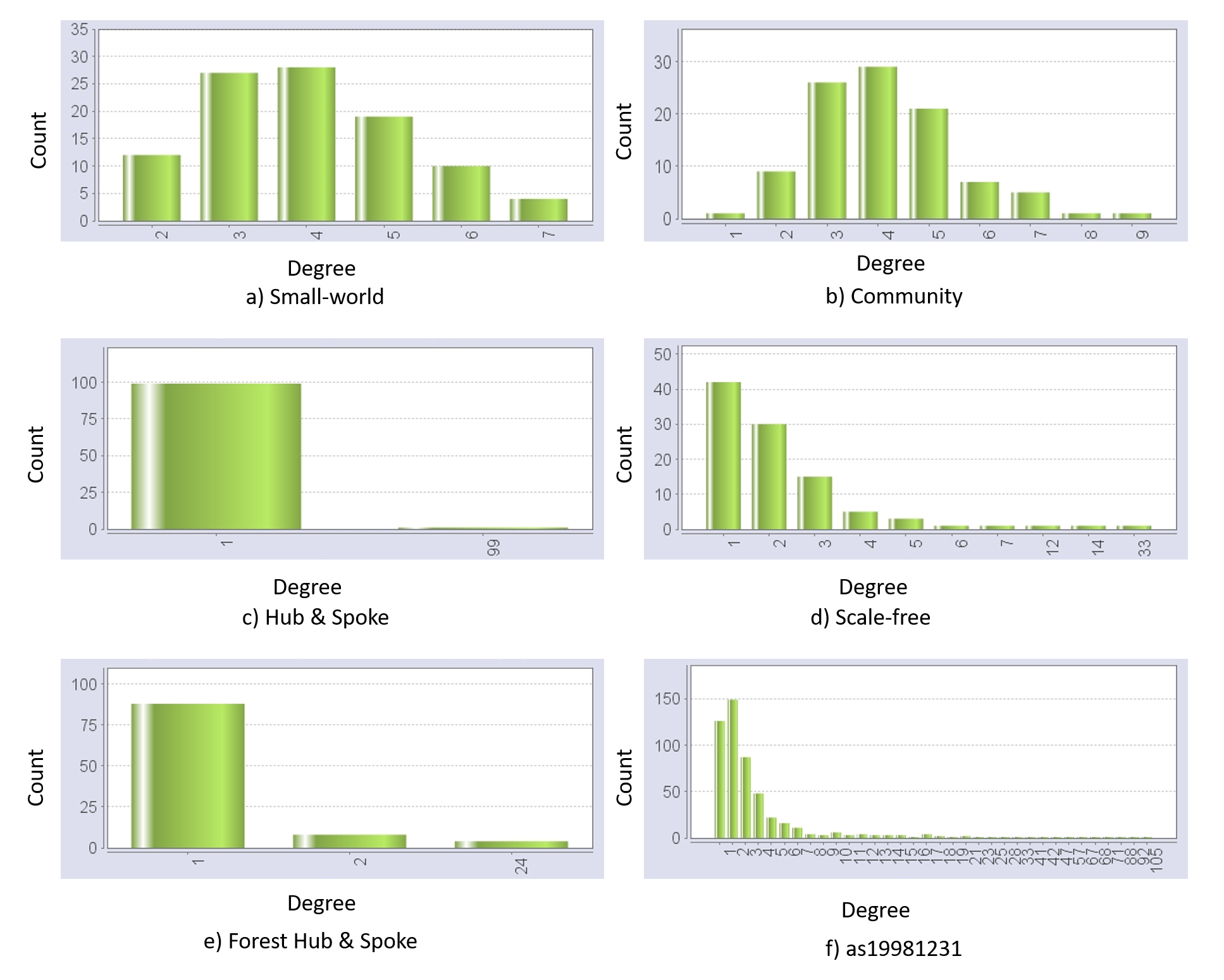}
\caption{Degree distributions for the selected network topologies}
\label{fig:degreedist}       
\end{figure}

Figure \ref{fig:degreedist} presents the histograms of the node degree distributions for the selected topologies. {The node degree is presented on the x-axis and the number of nodes featuring this degree on the y-axis. The various shapes of the network degree distributions provide a qualitative means to differentiate between generic topology types. These, in turn, impact the efficiency of decentralised communication algorithms} and hence their self-healing abilities. 
{For instance, the network topology in Fig. \ref{fig:degreedist}-a) features a minimum node degree of 2 and Fig. \ref{fig:degreedist}-b) features only one node of degree 1. This topological characteristic makes these networks more robust to node failure, as most nodes have more than one neighbour that can detect and recover them if they failed. Comparatively, the networks in Fig. \ref{fig:degreedist}-c), d), e) and f) feature multiple nodes of degree 1, which have less chances of being recovered in case of failure.}
Table \ref{tbl:network-features} presents some of the differentiating features of the selected networks. 

\begin{table}
\centering
\caption{Main features of selected network topologies}
\label{tbl:network-features}
\begin{tabular}{|c|l|}
\hline
\textbf{Name} &
  \textbf{Network Features} \\ \hline
\textbf{AS19981231} &
  \begin{tabular}[c]{@{}l@{}}
  Number of vertices: 512\\  Number of edges: 1273\\ Average Clustering Coefficient: 0.246\\ Average degree: 4.793\\ Diameter: 8.0\\ stdev Betweeness centrality: 0.00314\end{tabular} \\ \hline
\textbf{Small-World} &
  \begin{tabular}[c]{@{}l@{}}Number of vertices: 100\\  Number of edges: 200\\ Average Clustering Coefficient: 0.08014\\ Average degree: 4.0\\ Diameter: 6.0\\ stdev Betweeness centrality: 0.0048\end{tabular} \\ \hline
\textbf{Community} &
  \begin{tabular}[c]{@{}l@{}}Number of vertices: 100\\  Number of edges: 206\\ Average Clustering Coefficient: 0.132\\ Average degree: 4.12\\ Diameter: 11.0\\ stdev Betweeness centrality: 0.0174\end{tabular} \\ \hline
\textbf{\begin{tabular}[c]{@{}c@{}}Forest Hub\\  \& Spoke\end{tabular}} &
  \begin{tabular}[c]{@{}l@{}}Number of vertices: 100\\  Number of edges: 100\\ Average Clustering Coefficient: 0.0\\ Average degree: 2.0\\ Diameter: 8.0\\ stdev Betweeness centrality: 0.11153\end{tabular} \\ \hline
\textbf{Hub \& Spoke} &
  \begin{tabular}[c]{@{}l@{}}
  Number of vertices: 100\\  Number of edges: 99\\
  Average Clustering Coefficient: 0.0\\ Average degree: 1.98\\ Diameter: 2.0\\ stdev Betweeness centrality: 0.10206\end{tabular} \\ \hline
\textbf{Scale-free} &
  \begin{tabular}[c]{@{}l@{}}Number of vertices: 100\\  Number of edges: 127\\ Average Clustering Coefficient:  0.2509\\ Average degree: 2.54\\ Diameter: 9.0\\ stdev Betweeness centrality: 0.1392\end{tabular} \\ \hline
\end{tabular}
\end{table}

\subsubsubsection{Nodes}

A node is a static agent representing one process (or station) in the selected network. When a simulation begins, each node is initialised as follows. The node's communication buffer is empty; its memory contains topological information that depends on each experiment (Cf. sec \ref{subsec:exp-overview}); and, the node has the static agent program defined in Algorithm \ref{alg:basicNodeProgram}. Finally, the node's failure probability $p_f$ is set for the entire experiment. Values tested across experiments include $p_f= 0.5, 0.25, 0$.  

To evaluate the limits of the network recovery algorithm for increasing failure rates, we ran a set of experiments where the entire network's topological information was provided to all nodes, in their local memory, at initialisation time. The other experiments used either Trickle or the Mobile Agents-based process to collect topological data and store it into the nodes. For these experiments, each node's local memory was initialised with information on the node's direct neighbours only (needed to propagate data and heartbeat signals).  

\subsubsubsection{Mobile Agents}

Each mobile agent is provided with: a memory to store topology data obtained from nodes, a movement algorithm, and an agent program to communicate with visited nodes, simulate failures, and select its next movement. 

\section{Sub-models}

\subsection{The node failure model}

Nodes fail with a probability $p_f$, modelling node crashes in real distributed systems. At each simulation round, each node generates a uniform random number within [0,1]. If the random number is lower than $p_f$, then the node fails; all the agents residing on the node are removed; and connections to neighbouring nodes are deleted (Cf. lines 2-5 in Algorithm \ref{alg:basicNodeProgram}). 
\begin{algorithm}
\begin{scriptsize}
\caption{Basic Node Program}
\label{alg:basicNodeProgram}
\begin{algorithmic}[1]
\Procedure{act}{$node$}
    \State $rnd = getRandomNumber \in [0,1]$
    \If{$rnd < p_f \land isFailurePeriod()$} \Comment{Makes a node fail in a round interval}
        \State killNode($node$)
    \EndIf
    \State node.processMessages() \Comment{Send or receive data depending of data collection algorithm}
    \State node.evaluateNodeCreation() \Comment{Evaluates the creation of missing nodes}
\EndProcedure
\end{algorithmic}
\end{scriptsize}
\end{algorithm}

\subsection{Data-collection and dissemination model}

At each round, nodes collect and disseminate data via the $processMessages()$ method (line 6 in Algorithm \ref{alg:basicNodeProgram}). This method's definition
depends on the simulation's data-collection mode: (case-1) all topological information is provided to the nodes initially; (case-2) topological information is collected via Trickle; or (case-3) collected via Mobile Agents. 

If all nodes detain initial information about the complete network topology (case-1), then method $processMessages$ is defined in Algorithm \ref{alg:process-all-info}. This algorithm corresponds to a part of the recovery algorithm that connects a node with another node. When a node connects to a newly created replica (lines 3-5), it shares its local topology data with the new node, by sending a $networkData$ message.

\begin{algorithm}
\begin{scriptsize}
\caption{Process Messages if nodes have all the network information}
\label{alg:process-all-info}
\begin{algorithmic}[1]
\Procedure{processMessages}{$node$}
    \ForEach{$message \in node.MessagesQueue$}
        \If{$message[0] = connect$} \Comment{connect includes the node to connect id}
          \State connect(nodeToConnect) \Comment{connects to node to connect}
          \State $sendMessage(nodeToConnect, networkData \vert
          node.topologyDataFromMemory$ 
        \EndIf
        \If{$message[0] = networkData$}
            \State $union = node.topologyDataFromMemory \cup message[1]$ \Comment{$message[1]$ contains topology data received}
            \State $node.topologyDataFromMemory = union$
        \EndIf
    \EndFor
\EndProcedure
\end{algorithmic}
\end{scriptsize}
\end{algorithm}

When using Trickle for runtime data-collection,  $processMessages$ is defined by Algorithm \ref{alg:process-messages-trickle}.
{For the purpose of this paper, Trickle is adjusted to run in discrete rounds and included in the node process' messages function, with an adaptation of the original rules defined in RFC 6206 \cite{levis2011trickle}. In the experiments, Trickle's minimum interval size is set to $Imin = 1$ round, the maximum interval size is $Imax = 2^{16}$ rounds, and the redundancy constant is $k=3$. These values are recommended in the RFC6206, but translated here in rounds. $CurrentInterval$ is initialised as $Imin*2$ and the redundancy counter starts at $k=0$. $t$ is a random value obtained as shown at line 2.}
A node transmits its data to its neighbours at each $t$ rounds, unless its data is redundant with that already received from neighbouring nodes (line 3). 
To establish data redundancy, every time a node receives a message with topology data, it determines whether the received data is identical with its local topology data (line 12); and if so, it increases its redundancy $counter$. When the $currentInterval$ expires, the node reinitialises its redundancy counter ($counter=0$), and duplicates the value of $currentInterval$, until reaching a maximum value of $Imax$ (lines 32-39). Lines (23-30) define the part of the self-healing algorithm that shares the topology data when a node connects with a newly created node.

\begin{algorithm}
\begin{scriptsize}
\caption{Process Messages when using Trickle for runtime data-collection}
\label{alg:process-messages-trickle}
\begin{algorithmic}[1]
\Procedure{processMessages}{$node$}
    \State $t = randomNumber \in [currentInterval/2, currentInterval)$  \Comment{Set current interval size $I$}
    
    \If{$env.rounds \mod t = 0 \land counter < k $ } \Comment{Transmits at time t if the counter is lesser than the redundancy constant k} 
        \ForEach{$neigbour \in node.neighbours$} \Comment{Send topology data to neighbours}
            \State $sendMessage(neigbour, networkDataTrickle \vert
          node.topologyDataFromMemory$ 
        \EndFor
    \EndIf

    \ForEach{$message \in node.MessagesQueue$} \Comment{Process Incomming Messages}
        \If{$message[0] = networkDataTrickle$}
            \State $recvData = message[1]$
            \State $localData = node.topologyDataFromMemory$
            
            \If{$recvData \subset localData \land localData \subset recvData$} \Comment{If localData is equal that recvData}
                \State $counter \gets counter + 1$ \Comment{Consistent transmision detected}
            \Else
                \State $counter \gets 0$
                \State CurrentInterval = 1
                \State Imin = 1
                \State Imax = 1
            \EndIf
            
            \State $union = node.topologyDataFromMemory \cup message[1]$ \Comment{$message[1]$ contains topology data received}
            \State $node.topologyDataFromMemory = union$
        \EndIf
        \If{$message[0] = connect$} \Comment{connect includes the node to connect id}
          \State connect(nodetoConnect) \Comment{connects to node to connect}
          \State $sendMessage(nodetoConnect, networkData \vert
          node.topologyDataFromMemory$ 
        \EndIf
        \If{$message[0] = networkData$}
            \State $union = node.topologyDataFromMemory \cup message[1]$ \Comment{$message[1]$ contains topology data received}
            \State $node.topologyDataFromMemory = union$
        \EndIf
    \EndFor
    \If{$env.rounds \mod currentInterval == 0$} 
        \State $counter \gets 0$
        \State $currentInterval \gets 2*currentInterval$
        \If{$currentInterval > Imax$}
            \State $currentInterval = Imax$
        \EndIf
        \State Imin = $currentInterval/2$ 
        \State Imax = $currentInterval$
    \EndIf
\EndProcedure
\end{algorithmic}
\end{scriptsize}
\end{algorithm}

When using Mobile Agents for runtime data-collection (case-3), nodes receive topology information from the mobile agents that visit them. In this case, the $processMessages$ method is defined as in Algorithm \ref{alg:process-messages-mobile-agents}. Nodes receive topology data via messages of type  $networkDataMobileAgent$, and they store the union between their local data and the received data (lines 4-7). The remaining part of the algorithm includes the messages for connecting to another node, as necessary for the network recovery algorithm. 

\begin{algorithm}
\begin{scriptsize}
\caption{Process Messages when using Mobile Agents for data collection}
\label{alg:process-messages-mobile-agents}
\begin{algorithmic}[1]
\Procedure{processMessages}{$node$}
    \ForEach{$message \in node.MessagesQueue$} \Comment{Process Incomming Messages}
        \If{$message[0] = networkDataMobileAgent$}
            \State $recvData = message[1]$
            \State $localData = node.topologyDataFromMemory$
            \State $union = localData \cup recvData$ 
            \State $node.topologyDataFromMemory = union$
        \EndIf
        \If{$message[0] = connect$} \Comment{connect includes the node to connect id}
          \State connect(nodetoConnect) \Comment{connects to node to connect}
          \State $sendMessage(nodetoConnect, networkData \vert
          node.topologyDataFromMemory$ 
        \EndIf
        \If{$message[0] = networkData$}
            \State $union = node.topologyDataFromMemory \cup message[1]$ \Comment{$message[1]$ contains topology data received}
            \State $node.topologyDataFromMemory = union$
        \EndIf
    \EndFor
\EndProcedure
\end{algorithmic}
\end{scriptsize}
\end{algorithm}

\subsection{Mobile Agents movement, data management and failure model}
\label{subsubsubsec:mobile-agents-submodel}

The Mobile Agents program is defined in Algorithm \ref{alg:mobile-agent-program}. The program states that if the status of the agent's hosting node is $failed$ then the agent fails as well (line 3-5). Also, if the mobile agent's information is identical to the hosting node's information, then the agent increments its redundancy $counter$; otherwise 
the agent's $counter$ is reinitialised (lines 8-12). If the counter is equal to the redundancy factor $k$, then the mobile agent is discarded (lines 13-14). Finally, the mobile agent shares its collected data with its hosting node; updates its carried data  with topology data from its hosting node; and moves to the next node, following $motionAlgorithm$ (lines 16-20).

The agent movement algorithm is an implementation of a strategy developed previously \cite{Rodriguez2019} and referred to as \textit{Random with Marks} ($randomM$). When simulation starts, each node is marked as not visited ($visited=false$). Then, every time an agent passes through a node it marks the node as visited ($visited=true$). A mobile agent selects its next location via a uniformly distributed pseudo-random number from a collection of unmarked nodes ($visited=false$). This means that each agent must obtain the visited status of neighbouring nodes, with respect to its current location. As this is a boolean flag, it is counted as a ping in the estimation of the bandwidth overhead. In case that all neighbouring nodes are marked, one of the neighbours is selected randomly for the next move.

\begin{algorithm}
\begin{small}
\caption{Mobile Agent program}
\label{alg:mobile-agent-program}
\begin{algorithmic}[1]
\Procedure{act}{$mobileAgent$}
    \State Percept p \Comment{Collection with state of neighbours of currentLocation}
    \If{$currentLocation.status == Failed$}
        \State killMobileAgent($mobileAgent$)
    \EndIf
    \State $agentData = mobileAgent.topologyData$
    \State $nodeData = currentLocation.topologyDataFromMemory$
    \If{$agentData \subset nodeData \land nodeData \subset agentData$} 
        \State $counter \gets counter + 1$ \Comment{Consistent transmision detected}
    \Else
        \State $counter \gets 0$
    \EndIf
    \If{$counter == k $} \Comment{information is redundant for k consecutive steps}
        \State killMobileAgent($mobileAgent$)
    \EndIf
    \State $sendMessage(currentLocation, networkDataMobileAgent \vert agentData)$
    \State $mobileAgent.topologyData \gets nodeData \cup agentData$
    \State $currentLocation.visitedStatus \gets true$
    \State $Agent.move(motionAlgorithm(p))$
\EndProcedure
\end{algorithmic}
\end{small}
\end{algorithm}

\subsection{The self-healing model}
\label{subsubsec:self-healing}

Algorithm \ref{alg:rep} specifies how a node detects missing nodes based on the topology information obtained previously from external sources. Algorithm \ref{alg:nodecreation} presents the failure detection process as a local program in each node. 
{Only one neighbour of a node, connected to it directly, can recreate the node when it fails. Moreover, a neighbour can only recover a failed node if it has information about the failed node's neighbours. This allows neighbours to determine which one of them should actually recreate the missing node, based on the smallest id. As topological information spreads locally relatively fast, it is assumed that nearby neighbours have complete information about each other.} 

\begin{algorithm}[H]
\begin{scriptsize}
\caption{Evaluate node creation}
\label{alg:rep}
\begin{algorithmic}[1]
\Procedure{evaluateNodeCreation}{}
 	\State neighbours $\gets$ node.senseNeighbours(); \Comment{sense adjacent neighbours}	
 	\State memdata $\gets$ node.TopologyDataFromMemory; \Comment{gets data stored in memory}		
 	
 	\State $dif \gets ((neighbours \setminus memdata) \cup (memdata \setminus neighbours)$)
 	
	\If{$dif \neq \emptyset$} 
      \ForEach{$missingnode \in dif$}   \Comment{for each node id dif}		
			  \State neighDif = node.neighboursFromMemory(missingnode)  \Comment{Obtain the neighbourhood of the missing node from memory}
			  \If{$getMinimumId(neighDif) = node.id$} 
              		\State createNewNode(missingNode, neighDif)
              \EndIf
      \EndFor
	\EndIf
\EndProcedure
\end{algorithmic}
\end{scriptsize}
\end{algorithm}

{For instance, considering the network $a$--------$b$--------$c$, if node $b$ fails, then it  could be recreated by one of its neighbours -- either node $a$ or $c$. However, to do so, this neighbour must also have information about $b$'s neighbours. Considering that $a$ and $c$ hold the following information about $b$: ($b:[a,c]$), meaning that $b$ has $a$ and $c$ as neighbours, the neighbour with the minimum id -- $a$ in this case -- attempts to recreate $b$. Nodes with higher ids -- $c$ in this case -- will try to ping neighbours with lower ids -- $a$ here-- in case they have also failed. If a neighbour determines that nodes with lower ids fail to respond, it creates the missing node itself - e.g. if $c$ gets no reply from $a$ then it creates $b$. 
In the current implementation, the simulation runs in synchronous rounds and pings are executed via a unique dedicated channel. This ensures the creation of a single node for each id. This may also be the case, for instance, when creating networked devices, with unique IP addresses. However, in other systems, more complicated situations may occur due to parallel node executions -- such cases should be handled to ensure the uniqueness of node recreation (future work).}

In algorithm \ref{alg:nodecreation}, line 6 creates a new node with the same id of the missing node. If mobile agents are defined it also creates a new mobile agent in this location. Line 7 is important because a creator node shares its local topology information with the newly created node; hence allowing it to recreate missing neighbours, in turn. In lines 9-10 the creator node sends a message to the neighbours of a newly created node, making them connect to the new node, and share their local information with the new node. In the experiments performed, no redundant replicas of a recreated node were obtained when using these algorithms (\ref{alg:rep} and \ref{alg:nodecreation}).

\begin{algorithm}[H]
\begin{scriptsize}
\caption{Node Creation}
\label{alg:nodecreation}
\begin{algorithmic}[1]
\Procedure{CreateNewNode}{(missingnodeId, neighDif)}
 	\State $nod \gets findNode(missingnodeId)$; \Comment{Try to connect with the missing node before create}	 
    \If{$nod \neq null$}
    	\State connect(nod)  \Comment{Node reference already exists connecting instead create}	
    \Else
    	\State nod = createNewNode(missingId)
    \Comment{Creates a new node with id equal to missing node id. Creates also a mobile agent in this location if mobile agents are defined}
        \State nod.setNetworkdata(neighbours) \Comment{share topology information to new node}
        \State connect(nod)  \Comment{Node connect with recently created new node}	
        
        \ForEach{$neigh \in neighDif$} \Comment{request neighbours to connect with the new replica}
            \State sendMessage(neigh, $connect \vert nod.Id$) 
        \EndFor
    \EndIf 	
\EndProcedure
\end{algorithmic}
\end{scriptsize}
\end{algorithm}


\section{Experiments and Results}
\label{sec:experiments-and-results}

\subsection{Experimental considerations and settings}

{An experiment is defined by selecting: a) a network topology, b) a failure probability $p_f$, c) a data collection algorithm and d) a failure interval $(start, stop, end)$ defined as a period during which node failures occur (between $start$ and $stop$) followed by a recovery period until the $end$ round of simulation. Table \ref{tbl:parameters} summarises all parameters for each experiment.}

\begin{table}[H]
\caption{Parameters of experiments performed}
\label{tbl:parameters}
\begin{tabular}{|c|l|l|}
\hline
\multicolumn{3}{|c|}{\textbf{Experimental Parameters}}                                                                                   \\ \hline
network &
  initial network topology &
  \begin{tabular}[c]{@{}l@{}}as19981231, small-world, community\\ network, forest hub \& spoke, \\ hub \& spoke, scale-free.\end{tabular} \\ \hline
pf                        & failure probability       & defined in Table \ref{tbl:success-allinfo}\\ \hline
data collection algorithm & technique to collect data & Trickle, Mobile Agents                                                             \\ \hline
(start, stop, end) &
  \begin{tabular}[c]{@{}l@{}}Intervals of failure and end\\ of experiment\end{tabular} &
  \begin{tabular}[c]{@{}l@{}}Depends on the network used\\  in experiments with failures\end{tabular} \\ \hline
\end{tabular}
\end{table}

{For the Mobile Agents approach, the simulation invariants are: the number of mobile agents, which equals the number of nodes in each network, with one agent located in each node at the start; and, the agent movement algorithm, which was the same for all topologies (as shown to be the fastest one in previous work \cite{Rodriguez2019}).  Mobile agent parameters are defined in Table \ref{tbl:mobile-agent-parameters}. } 

\begin{table}[H]
\caption{Parameters of Mobile Agents}
\label{tbl:mobile-agent-parameters}
\centering
\begin{tabular}{|c|l|c|}
\hline
\multicolumn{3}{|c|}{\textbf{Mobile Agents Parameters}} \\ \hline
\textbf{Name} & \multicolumn{1}{c|}{\textbf{Meaning}} & \textbf{Initial Value}                                              \\ \hline
popSize       & number of agents                      & \multicolumn{1}{l|}{Same that number of nodes} \\ \hline
location      & initial location of agents            & \multicolumn{1}{l|}{Starts located in each node}     \\ \hline
Counter     & counter to estimate redundancy     & 0    \\ \hline
k           & redundancy constant                &3 (and 5 for AS network)    \\ \hline
\end{tabular}
\end{table}

{When Trickle is used parameters are initialised as in Table \ref{tbl:trickle-parameters}. The simulation invariants are the minimum and maximum interval sizes and the redundancy constant $k$.}

\begin{table}[H]
\caption{Parameters of Tricke}
\label{tbl:trickle-parameters}
\resizebox{\textwidth}{!}{%
\begin{tabular}{|c|l|r|}
\hline
\multicolumn{3}{|c|}{\textbf{Trickle Parameters}} \\ \hline
\textbf{Name} & \multicolumn{1}{c|}{\textbf{Meaning}} & \multicolumn{1}{c|}{\textbf{Initial Value}} \\ \hline
$Imin$ & minimum interval size & 1 round \\ \hline
$Imax$ & maximum interval size & 2e16 rounds \\ \hline
$k$ & redundancy constant & 3 (and 5 for AS network) \\ \hline
$currentInterval$ & current interval size & starts with 2 round and increases \\ \hline
$Counter$ & counter to estimate redundancy & 0 \\ \hline
$t$ & time to send data & \begin{tabular}[c]{@{}r@{}}random value in \\ $[currentInterval/2, currentInterval]$\end{tabular} \\ \hline
\end{tabular}}
\end{table}

{Experimental evaluation relies upon three main metrics, assessed at the end of each round:  \textit{memory usage}, in bytes -- for nodes, and, when applicable, for mobile agents; \textit{network similarity}, in percentages; and, \textit{bandwidth overhead}, in bytes and in number of messages received by nodes. 
Network similarity is determined by comparing the network topology in the current round with the original topology at the experiment's start (round zero). We adopted the graph similarity metric proposed in \cite{Nikolic2012} and implemented in \cite{Sashika2014}. This metric allows estimating the similarity between two graphs, generating a score within $[0,1]$. This score is calculated based on the principle that two graphs $A$ and $B$ are considered to be similar if, given two nodes $i \in nodes_A$ and $j \in nodes_B$, the neighbour nodes of $i$ can be matched to similar neighbours of $j$. In this work, the similarity value is calculated as a percentage -- e.g. 100\% similarity between isomorphic networks.}

{Memory consumption is calculated by obtaining the size (in bytes) of the topology data collected by all the agents (nodes and Mobile agents, separately) in each round. The bandwidth overhead corresponds to the total amount of bytes received by all the nodes in each round. The number of messages is the number of messages received by all the nodes in each round.
}

{
The following subsections introduce three kinds of experiments. In the first set of experiments, nodes are provided with the entire topology data of each selected network. These experiments determine the limits of the proposed self-healing algorithm with respect to the failure rate $pf$, for each network topology. 
A second set of experiments employ the two data-collection techniques, yet without node failures. It allows to determine the time required to obtain and disseminate the entire topology data in all the nodes. This, in turn, allows to set the start of the failure intervals in the third kind of experiments. This last experiment starts with an initial failure-free interval (determined previously), then undergoes a period where node failures occur, and finally allow for a recovery period before the simulation ends.} 

{\subsection{Experiments where nodes have all topological data: limits of $pf$}
\label{subsec:exp-overview}}

 The purpose {of these experiments} was to determine the limits of the recovery algorithm for increasing values of the failure probability $p_f$, for different topologies. All the nodes have initial information about the network topology and execute Algorithm \ref{alg:process-all-info} and Algorithm \ref{alg:rep} to recover the selected networks. Each experiment is repeated 30 times and runs for 100 rounds. Nodes fail with probability $p_f$ for the first 50 rounds, then stop failing for the remaining 50 rounds to determine whether the proposed approach can recover the original network. 
 
 Table \ref{tbl:success-allinfo}, presents the maximum $p_f$ values for which the tested networks recovered completely (i.e. 100\% similarity to the original topology) in round 100. Based on these results, each network topology was associated to a maximum $p_f$ value.

\begin{table}[H]
\caption{Network self-healing ability when nodes know the entire topology beforehand. The number of experiments that failed to recover the entire network is marked with $<Number>^{*}$, out of 30 experiments. Cases of full network recovery are marked with `Success'.}
\label{tbl:success-allinfo}
\centering  
\begin{tabular}{|l|c|c|}
\hline
\multicolumn{1}{|c|}{\multirow{2}{*}{Network}} & \multicolumn{2}{c|}{pf} \\ \cline{2-3} 
\multicolumn{1}{|c|}{}                         & 0.5        & 0.25       \\ \hline
as19981231                                     & Success    &            \\ \hline
Community Network                              & Success    &            \\ \hline
Small-World                                    & Success    &            \\ \hline
Scale-free                                     & 6*         & Success    \\ \hline
Forest Hub and Spoke                           & 7*         & Success    \\ \hline
Hub \& Spoke                                   & 8*         & Success    \\ \hline
\end{tabular}
\end{table}


{Once the maximum $p_f$ value for self-healing each topology was established for the case where the topology was known in advance, the remaining problem was how to enable nodes to obtain this topology information at runtime.}

{\subsection{Experiments to determine reference time points for collecting the entire topology data}\label{subsec:exp-no-failures}}

A second set of experiments were performed {with $pf=0$} to determine the number of rounds required for the two data-collection approaches to obtain the entire topology information. Experiments were performed for the adapted Trickle algorithm (Cf. Algorithm \ref{alg:process-messages-trickle}), and for the proposed Mobile Agents approach (Cf. Algorithms  \ref{alg:process-messages-mobile-agents} and \ref{alg:mobile-agent-program}). The results were used as time references for introducing node failures in subsequent experiments. 
 
\begin{figure}[H]
\centering
\includegraphics[width=\textwidth]{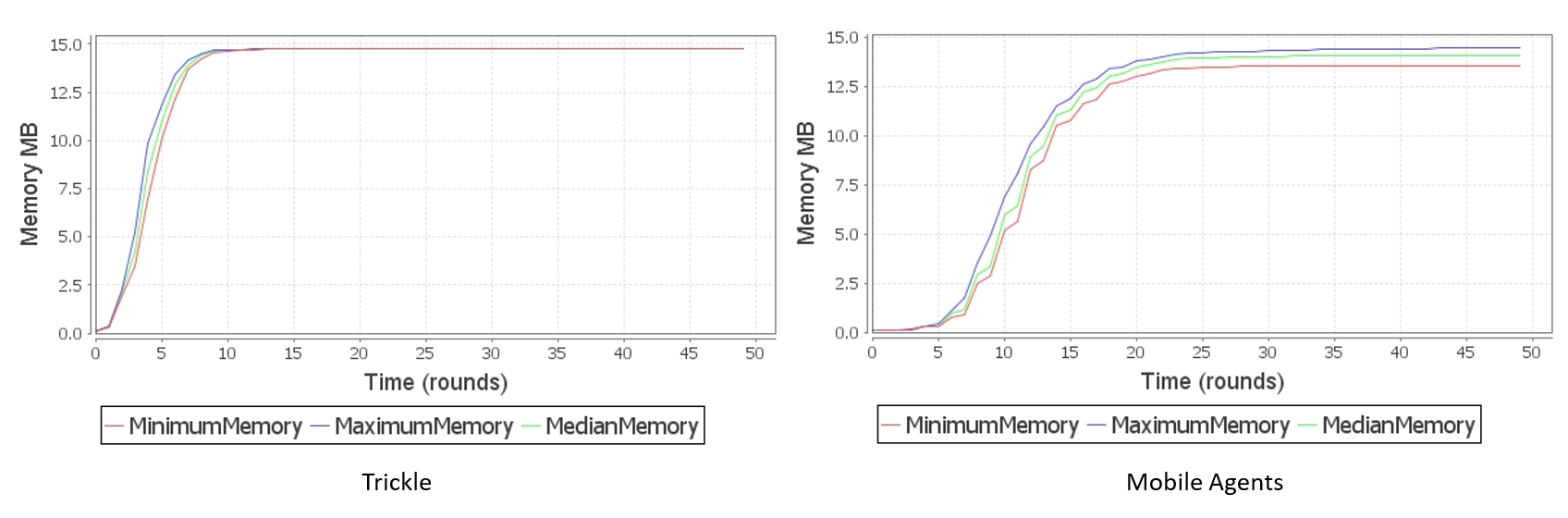}
\caption{Round number vs Data Collected by nodes (in Megabytes), for the AS network}
\label{fig:memorynofail}       
\end{figure}

{Figure \ref{fig:memorynofail} shows how the amount of collected data increases over time (in rounds) when Trickle and Mobile Agents are employed, for the AS network (as1998123). The graph shows data measurements in Megabytes, including the  minimum, median and maximum values over the 30 experimental runs.} In all failure-free experiments, Trickle collected and disseminated the complete network topology data to all nodes. After several initial rounds, the minimum, median and maximum size of data collected reaches the same value, which corresponds to the representation size of the entire topology.

When Mobile Agents are used, results show that agents collect and disseminate topology data only partially, because the minimum, median and maximum memory values are different and below the maximum value (i.e. the size of complete topology representation). At a certain point, the amount of collected data reaches a plateau (no more changes). This is the point where agents start being deleted due to the redundancy rule defined in the mobile agent’s program (i.e. they no longer provide novel data to visited nodes). Results also show that Trickle collects topology data faster than Mobile Agents. This can be expected considering Trickle's gossip-based approach (broadcast) versus the Mobile Agents' limited exploration (Random with Markings).

Table \ref{tbl:pointsref} presents the selected reference points (in rounds),  when the amount of data collected reaches a static value; suggesting that the maximum topology information was collected.

\begin{table}[H]
\caption{Points of reference for experiments with failures}
\label{tbl:pointsref}
\centering    
\begin{tabular}{l|c|c|}
\cline{2-3}
                                                    & \textbf{Trickle} & \textbf{Mobile Agents} \\ \hline
\multicolumn{1}{|l|}{\textbf{AS}}                   & 13               & 38                     \\ \hline
\multicolumn{1}{|l|}{\textbf{Small-World}}          & 12               & 25                     \\ \hline
\multicolumn{1}{|l|}{\textbf{Community Network}}    & 20               & 41                     \\ \hline
\multicolumn{1}{|l|}{\textbf{Forest Hub and Spoke}} & 17               & 35                     \\ \hline
\multicolumn{1}{|l|}{\textbf{Hub \& Spoke}}         & 5                & 14                     \\ \hline
\multicolumn{1}{|l|}{\textbf{Scale-free}}           & 17               & 33                     \\ \hline
\end{tabular}
\end{table}

The reference points were selected based on experimental results such as those exemplified in Figure \ref{fig:memorynofail} for the as1998123 network. For Trickle, the reference point is easy to choose because the minimum, median and maximum values are quasi-identical (i.e. 13 rounds in Fig.  \ref{fig:memorynofail}).  For Mobile Agents, the reference point was selected based on the median round at which the number of agents drops to zero, for each network. It is observed that no further data is collected after this point (e.g. round 38 in Fig.  \ref{fig:memorynofail}). The reference points obtained for both data-collection algorithms, for each network topology, were used as starting points for introducing node failures in subsequent experiments.

{\subsection{Experiments with node failures, using the reference time points}}
\label{subsec:exp-with-failures}

Self-healing experiments were configured to run without failure until the reference points established previously (Table \ref{tbl:pointsref}); then {to} fail for 25 rounds (with $p_f$); and finally {to} run without failure again for another 25 rounds. Within each experiment, these reference points are marked as `start' (nodes start failing), `stop' (nodes stop failing) and `end' (end of the experiment). The two reference points obtained for Trickle and Mobile Agents were tested in each experiment, for both approaches. As Trickle collected data faster than the Mobile Agents (e.g. 13 rounds vs 38 rounds for the `AS' network), using Mobile Agents while introducing failures at Trickle's `start' reference point (e.g. round 13 for `AS') meant that the Mobile Agents approach had to recover the network even if they had not yet collected the maximum topology information. 

For instance, for the $as19981231$ network (`AS'), two experiments were performed for each data-collection approach (Trickle and {Mobile}-Agents): i) one where nodes started failing at round 13, stopped failing at round 38, and ended at round 63 -- refereed to as `(13, 68, 63)' in the reported results; and, ii) another one that started failing at round 38, stopped failing at round 63, and ended at round 88 -- referred to as `(38, 63, 88)' in the reported results. The tested failure probability value $p_f$ was determined from the experiments where all the information was provided (Table \ref{tbl:success-allinfo}).

\begin{figure}[H]
\centering
\includegraphics[width=\textwidth]{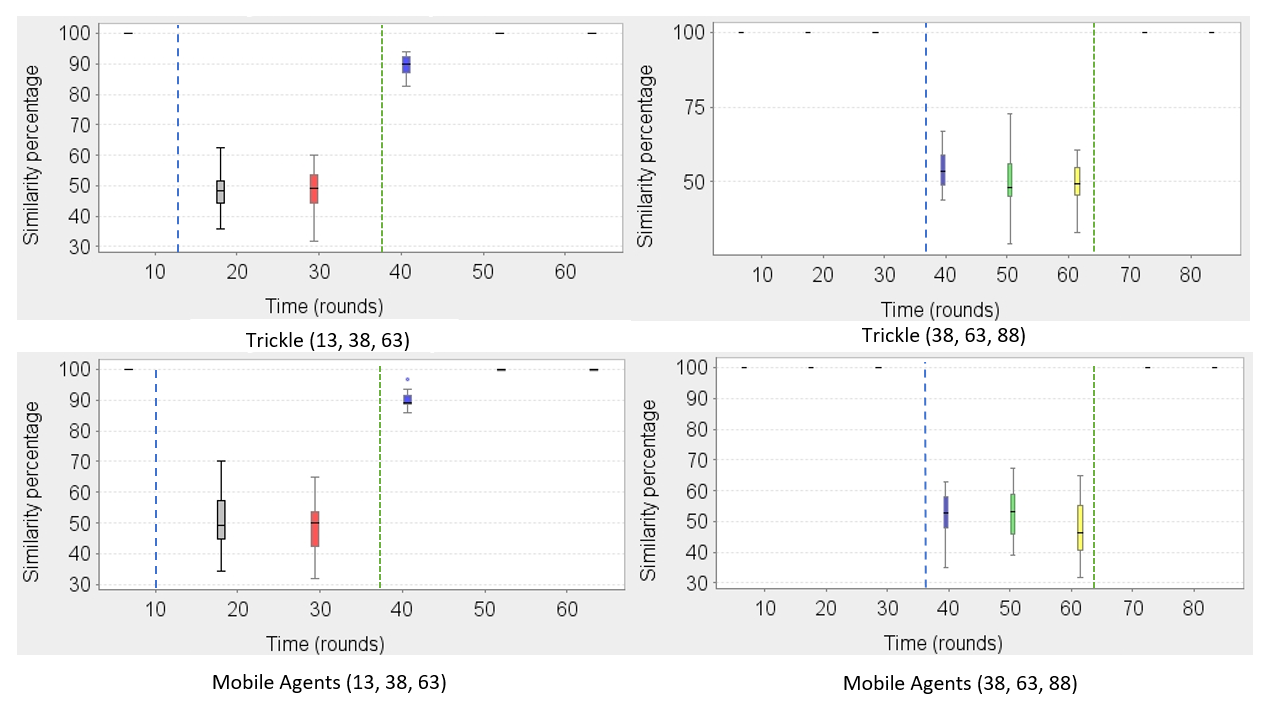}
\caption{Round number vs similarity for AS network and pf = 0.5, k = 3}
\label{fig:similaritiesfail}       
\end{figure}

Figure \ref{fig:similaritiesfail} shows the network similarity over time (in rounds), for Trickle and {Mobile} Agents, for the AS network. In each case, we tested two scenarios, each one starting to introduce node failures after a different round number -- corresponding to the reference points at which Trickle and Multi Agents managed to collect and disseminate a maximum of topological information {(Cf. subsubsec. \ref{subsec:exp-no-failures}).}  {In Figure \ref{fig:similaritiesfail}}, the blue dotted line indicates the round number when the nodes \textit{start} to fail; and the green dotted lines the point when the system \textit{stops} failing. For both Trickle and Mobile Agents, network similarity drops during the node-failure period and recovers to 100\% (or almost 100\%, Cf. below) after the failures period expires.

In all experiments using the reference points of Mobile Agents (i.e. the time needed by agents to disseminate topology information to all nodes) both Trickle and Mobile Agents managed to recover the network entirely, for all network topologies. This implies that even if Mobile Agents only disseminate partial topological information into nodes, this suffices for completely recovering all the selected networks; as if the nodes stored all the topology information. This is probably due to the fact that the nodes receive mostly information about their neighbourhood, which is the important one for recovering the local topology. The information that the nodes are missing is most likely about remote nodes, which they do not have to repair.  

\begin{figure}[H]
\centering
\includegraphics[scale=0.45]{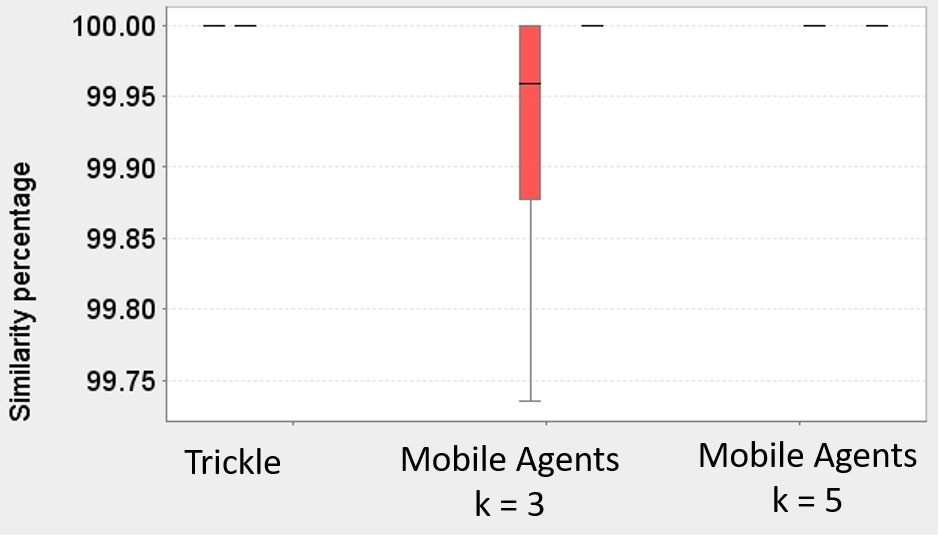}
\caption{Compiled Similarity AS network and pf = 0.5}
\label{fig:compsimilaritiesfail}       
\end{figure}

{When using Trickle's reference time points, for all network topologies except for the AS network, the same redundancy factor $k=3$ suffices for both Trickle and Mobile Agents to recover the entire network. For the AS network,} {Figure \ref{fig:compsimilaritiesfail} presents the similarity percentages for each data collection technique with respect to the two points of reference of the AS Network --Trickle $(13, 68, 63)$ and Mobile Agents $(38, 63, 88)$. It is observed that for Trickle and for Mobile Agents with $k=5$ the similarity percentage is 100\%. However, for the AS network, using Mobile Agents with a redundancy factor of $k=3$ and the Trickle reference point, resulted in a median similarity of 99.95\% (Cf. Figure \ref{fig:compsimilaritiesfail})}.

\begin{figure}[H]
\centering
\includegraphics[width=\textwidth]{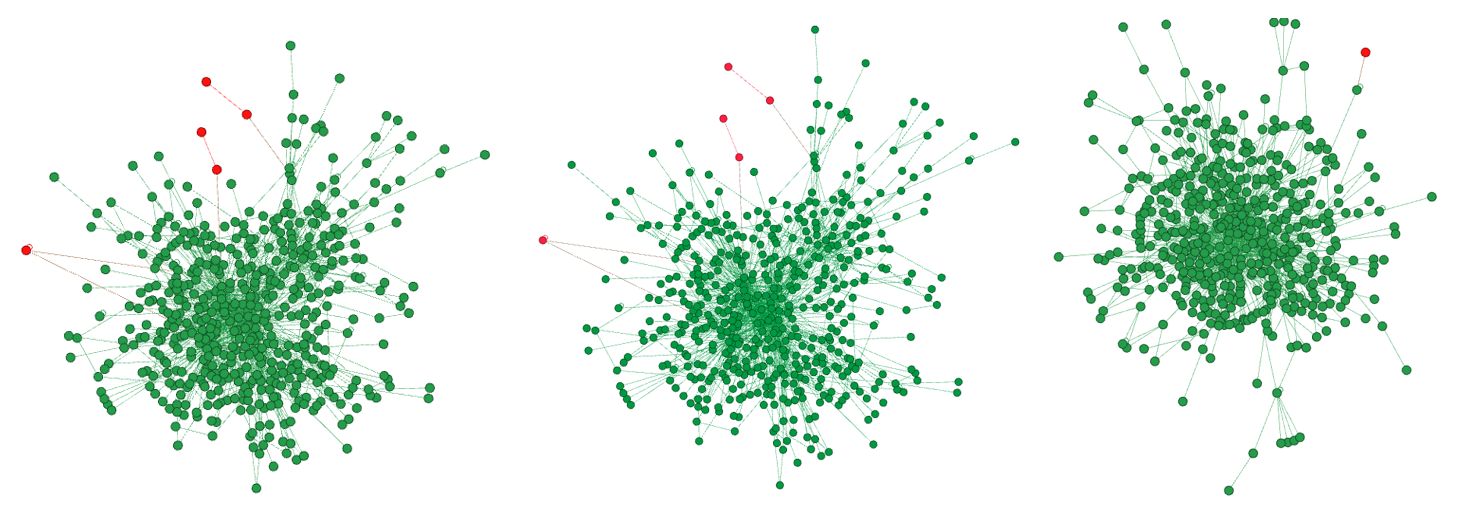}
\caption{Three examples of topology snapshots, for the AS network, at the end of each experiment. Green nodes are those that are alive / recovered. Red nodes are those missing with respect to the original topology. {Experiments used Mobile Agents with $k=3$ and Trickle reference points $(13, 68, 63)$.}}
\label{fig:missingnodes}       
\end{figure}

To establish the most sensitive nodes (i.e. most likely to be lost), a comparison was performed between the network topology at the beginning and at the end of the experiment (i.e. before node failures and after the node recovery period). This indicated that the lost nodes featured degree two at most, with one of their neighbours of degree one. {Figure \ref{fig:missingnodes} corresponds to three experiments performed with the AS network (with $k=3$, Trickle's points of time reference and Mobile Agents as data-collection technique). Missing nodes at the end of each experiment appear in red.} {As in Figure \ref{fig:missingnodes}, nodes of degree one are more likely to be lost as a single neighbour can recover them; if this neighbour is also lost, or does not get the topological information in time, then the marginal node is lost. Similarly, for nodes of degree two, once their neighbour of degree one fails without being recovered, the node itself becomes of degree one, hence more prone to oblivion.}

{Based on these preliminary results in Figures \ref{fig:compsimilaritiesfail} and  \ref{fig:missingnodes}, for the AS network, we re-calibrated the simulation and increased the redundancy factor to $k=5$ for the Mobile Agents. This allowed collecting more information and recovering the AS network with 100\% success rate. Hence, resource overheads incurred for the Mobile Agents approach were measured for the case where the redundancy factor was set to  $k=5$.}
\\

\subsection{Resource overheads}

\begin{figure}[H]
\centering
\includegraphics[scale=0.45]{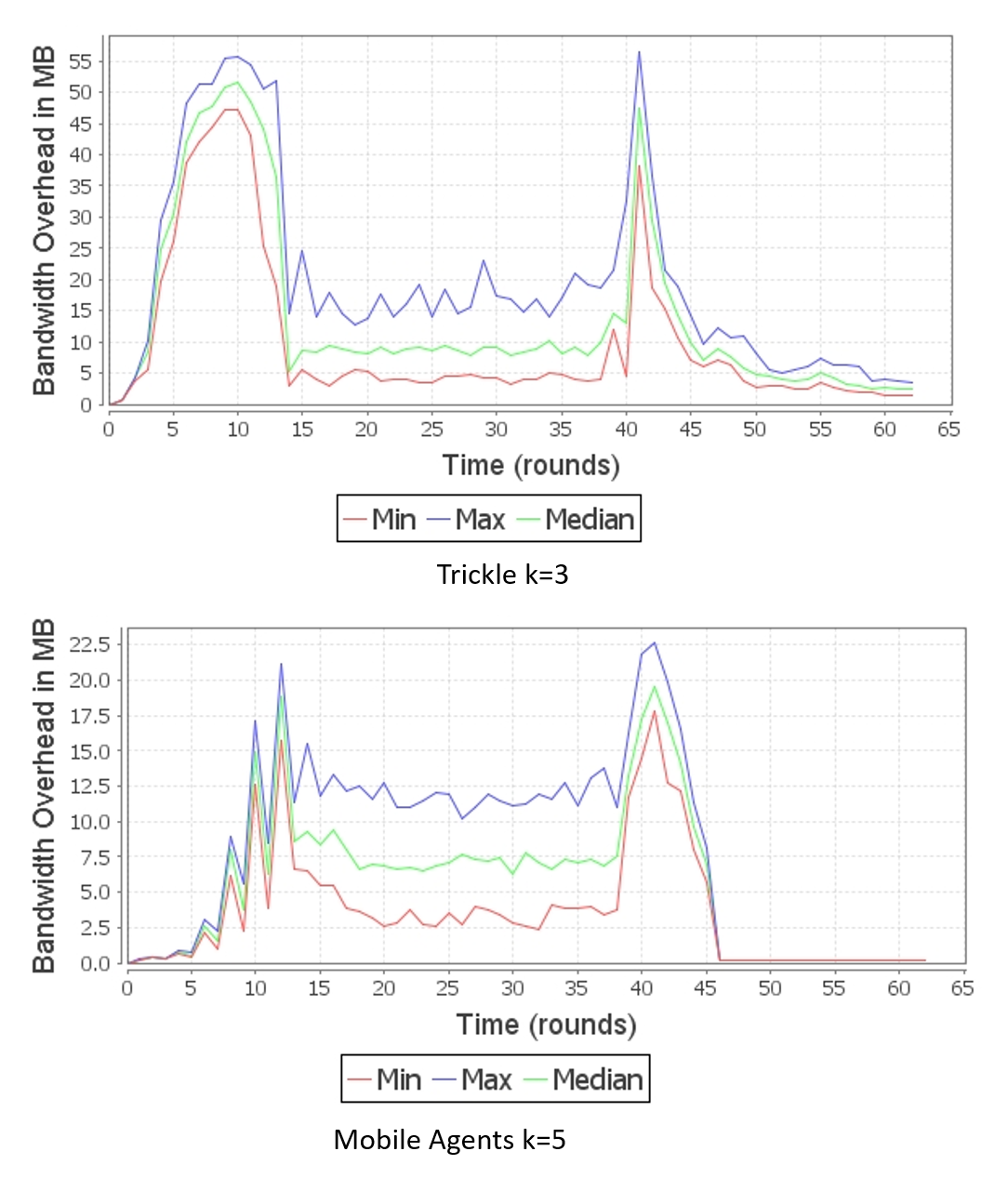}
\caption{Bandwidth overhead in Megabytes for AS pf=0.5 (13, 38, 63)}
\label{fig:bwoverhead}       
\end{figure}

Figure \ref{fig:bwoverhead} shows the amount of data transferred between all nodes in the AS network, at every round, in Megabytes. The depicted experimental scenarios ran for 13 rounds without failure (to collect topology data), then started to fail until round 38, and finally allowed nodes to recover until the end (round 63) -- i.e. experiment (13, 38, 63).
{In Figure \ref{fig:bwoverhead} it is observed that in the AS network, Mobile Agents incurred  less bandwidth overheads, even with a k=5, compared with Trickle that used a lesser redundancy factor of k=3.}  

{In Figure \ref{fig:bwoverhead}, it is also observed that before round 13, in the data collection interval without failures, Mobile Agents carry less information over the network than Trickle. The period between round 13 and 38 introduces node failures with a $pf = 0.5$, causing a significant amount of nodes to fail. After round 38 failures stop and it is observed that bandwidth is increased because missing nodes are being recreated. Once these missing nodes are created the bandwidth consumption is reduced for both Trickle and Mobile Agents. For Mobile Agents bandwidth consumption is reduced to zero because after some time agents that carry redundant data are dropped. For Trickle, bandwidth consumption is reduced close to zero because transmission redundancy causes an increase in the next transmission interval size and stops sending redundant data once $k$ is reached. A similar behaviour can be observed for the other networks.}

Table \ref{tbl:bw-overhead} {summarises the amounts of total communication overheads for all the reference points  (e.g., (13, 38, 63) and (38, 63, 88)), with different redundancy factors (e.g. $k=3$, $k=5$), for all selected topologies, when using Trickle and Mobile Agents approaches. Table \ref{tbl:bw-overhead} includes the
integral over the entire experiment 
of the minimum, median and maximum size of messages received by nodes for all the selected networks during the experiments.} {In Figure \ref{fig:bwoverhead} and Table \ref{tbl:bw-overhead} it is observed that the median and maximum amount of byte transfers is smaller for Mobile Agents than for Trickle. This is because agents carry less data than Trickle messages do. Also, agents incur smaller messages between nodes (not considering ping messages which are negligible).}

\begin{table}[H]
\centering
\caption{Bandwidth overheads in bytes. The redundancy factor for Trickle was $k=3$}
\label{tbl:bw-overhead}
\resizebox{\textwidth}{!}{%
\begin{tabular}{|l|c|c|c|c|c|c|c|c|c|c|c|}
\hline
\multicolumn{1}{|c|}{\multirow{2}{*}{\textbf{Network}}} &
  \multirow{2}{*}{\textbf{\begin{tabular}[c]{@{}c@{}}k for mobile\\ agents\end{tabular}}} &
  \multirow{2}{*}{\textbf{\begin{tabular}[c]{@{}c@{}}Exp. Config.\end{tabular}}} &
  \multicolumn{3}{c|}{\textbf{Trickle}} &
  \multicolumn{3}{c|}{\textbf{Mobile Agents}} &
  \multicolumn{3}{c|}{\textbf{Difference Percentage RPD}} \\ \cline{4-12} 
\multicolumn{1}{|c|}{} &
   &
   &
  \textbf{Max} &
  \textbf{Median} &
  \textbf{Min} &
  \textbf{Max} &
  \textbf{Median} &
  \textbf{Min} &
  \textbf{Max} &
  \textbf{Median} &
  \textbf{Min} \\ \hline
\multirow{4}{*}{\textbf{AS19981231}} &
  3 &
  13, 38, 83 &
  6.24E+09 &
  4.41E+09 &
  3.19E+09 &
  2.30E+09 &
  1.56E+09 &
  9.83E+08 &
  92.31\% &
  95.45\% &
  105.63\% \\ \cline{2-12} 
 &
  3 &
  38, 63, 88 &
  7.17E+09 &
  5.04E+09 &
  3.60E+09 &
  3.46E+09 &
  2.36E+09 &
  1.51E+09 &
  69.92\% &
  72.44\% &
  81.75\% \\ \cline{2-12} 
 &
  5 &
  13, 38, 63 &
  6.24E+09 &
  4.41E+09 &
  3.19E+09 &
  2.44E+09 &
  1.71E+09 &
  1.11E+09 &
  87.66\% &
  88.44\% &
  96.58\% \\ \cline{2-12} 
 &
  5 &
  38, 63, 88 &
  7.17E+09 &
  5.04E+09 &
  3.60E+09 &
  3.90E+09 &
  2.65E+09 &
  1.71E+09 &
  59.16\% &
  62.11\% &
  71.23\% \\ \hline
\multirow{2}{*}{\textbf{Small-World}} &
  3 &
  12, 37, 62 &
  1.60E+08 &
  1.23E+08 &
  8.89E+07 &
  7.40E+07 &
  5.59E+07 &
  3.89E+07 &
  73.25\% &
  74.76\% &
  78.33\% \\ \cline{2-12} 
 &
  3 &
  25, 50, 75 &
  1.89E+08 &
  1.42E+08 &
  1.04E+08 &
  9.90E+07 &
  7.59E+07 &
  5.73E+07 &
  62.51\% &
  60.97\% &
  57.80\% \\ \hline
\multirow{2}{*}{\textbf{Community}} &
  3 &
  20, 45, 60 &
  2.03E+08 &
  1.52E+08 &
  1.10E+08 &
  8.57E+07 &
  6.07E+07 &
  3.91E+07 &
  81.06\% &
  85.97\% &
  94.90\% \\ \cline{2-12} 
 &
  3 &
  41, 66, 91 &
  2.33E+08 &
  1.72E+08 &
  1.20E+08 &
  9.88E+07 &
  7.05E+07 &
  4.68E+07 &
  81.03\% &
  83.86\% &
  87.79\% \\ \hline
\multirow{2}{*}{\textbf{\begin{tabular}[c]{@{}l@{}}Forest Hub\\  \& Spoke\end{tabular}}} &
  3 &
  17, 42, 67 &
  1.14E+08 &
  6.88E+07 &
  3.66E+07 &
  6.41E+07 &
  4.03E+07 &
  2.14E+07 &
  55.69\% &
  52.27\% &
  52.48\% \\ \cline{2-12} 
 &
  3 &
  35, 60, 85 &
  1.22E+08 &
  7.39E+07 &
  3.90E+07 &
  7.13E+07 &
  4.29E+07 &
  2.35E+07 &
  52.70\% &
  53.20\% &
  49.78\% \\ \hline
\multirow{2}{*}{\textbf{Scale-free}} &
  3 &
  17, 42, 67 &
  1.40E+08 &
  9.44E+07 &
  5.81E+07 &
  7.05E+07 &
  4.77E+07 &
  3.04E+07 &
  66.00\% &
  65.76\% &
  62.59\% \\ \cline{2-12} 
 &
  3 &
  33, 58, 83 &
  1.54E+08 &
  1.02E+08 &
  6.45E+07 &
  7.70E+07 &
  5.34E+07 &
  3.43E+07 &
  66.85\% &
  62.75\% &
  61.27\% \\ \hline
\multirow{2}{*}{\textbf{\begin{tabular}[c]{@{}l@{}}Hub \& \\ Spoke\end{tabular}}} &
  3 &
  5, 30, 55 &
  1.19E+08 &
  5.30E+07 &
  1.26E+07 &
  7.27E+07 &
  4.39E+07 &
  1.34E+07 &
  48.71\% &
  18.73\% &
  6.30\% \\ \cline{2-12} 
 &
  3 &
  14, 39, 64 &
  1.25E+08 &
  5.96E+07 &
  1.66E+07 &
  8.34E+07 &
  5.01E+07 &
  1.84E+07 &
  39.84\% &
  17.20\% &
  10.24\% \\ \hline
\end{tabular}%
}
\end{table}

In terms of number of messages, Trickle exchanges fewer messages, in total, than Mobile Agents, in all the experiments performed. {Table \ref{tbl:number-messages}, presents the integral over the entire experiment of the number of messages received by nodes (minimum, median and maximum)}). However, Mobile Agents incur fewer large messages (i.e. carrying topology information) between nodes. While Trickle uses gossip propagation (broadcasts from each node), for Mobile Agents, the number of data-collection messages in each round is limited to the number of available agents. 

\begin{table}[H]
\centering
\caption{Number of messages received by all nodes. Redundancy factor for Trickle: $k=3$}
\label{tbl:number-messages}
\resizebox{\textwidth}{!}{%
\begin{tabular}{|l|c|c|c|c|c|c|c|c|c|c|c|}
\hline
\multicolumn{1}{|c|}{\multirow{2}{*}{\textbf{Network}}} &
  \multirow{2}{*}{\textbf{\begin{tabular}[c]{@{}c@{}}k for mobile\\  agents\end{tabular}}} &
  \multirow{2}{*}{\textbf{\begin{tabular}[c]{@{}c@{}}Exp.\\ Config.\end{tabular}}} &
  \multicolumn{3}{c|}{\textbf{Trickle}} &
  \multicolumn{3}{c|}{\textbf{Mobile Agents}} &
  \multicolumn{3}{c|}{\textbf{Difference Percentage RPD}} \\ \cline{4-12} 
\multicolumn{1}{|c|}{} &
   &
   &
  \multicolumn{1}{l|}{\textbf{Max}} &
  \multicolumn{1}{l|}{\textbf{Median}} &
  \multicolumn{1}{l|}{\textbf{Min}} &
  \multicolumn{1}{l|}{\textbf{Max}} &
  \multicolumn{1}{l|}{\textbf{Median}} &
  \multicolumn{1}{l|}{\textbf{Min}} &
  \multicolumn{1}{l|}{\textbf{Max}} &
  \multicolumn{1}{l|}{\textbf{Median}} &
  \multicolumn{1}{l|}{\textbf{Min}} \\ \hline
\multirow{4}{*}{\textbf{AS19981231}}           & 3 & 13, 38, 63 & 1.74E+06 & 1.25E+06 & 8.59E+05 & 2.97E+06 & 2.29E+06 & 1.65E+06 & 101.53\% & 103.22\% & 108.06\% \\ \cline{2-12} 
                                               & 3 & 38, 63, 88 & 2.09E+06 & 1.59E+06 & 1.16E+06 & 4.12E+06 & 3.08E+06 & 2.21E+06 & 33.52\%  & 36.71\%  & 36.58\%  \\ \cline{2-12} 
                                               & 5 & 13, 38, 63 & 1.74E+06 & 1.25E+06 & 8.59E+05 & 3.14E+06 & 2.45E+06 & 1.78E+06 & 57.48\%  & 64.42\%  & 69.70\%  \\ \cline{2-12} 
                                               & 5 & 38, 63, 88 & 2.09E+06 & 1.59E+06 & 1.16E+06 & 4.52E+06 & 3.35E+06 & 2.38E+06 & 73.32\%  & 71.19\%  & 68.69\%  \\ \hline
\multirow{2}{*}{\textbf{Small-World}}          & 3 & 12, 37, 62 & 1.62E+05 & 1.37E+05 & 1.15E+05 & 1.97E+05 & 1.61E+05 & 1.30E+05 & 19.03\%  & 16.01\%  & 12.31\%  \\ \cline{2-12} 
                                               & 3 & 25, 50, 75 & 1.95E+05 & 1.66E+05 & 1.42E+05 & 2.49E+05 & 2.13E+05 & 1.86E+05 & 24.17\%  & 24.58\%  & 26.45\%  \\ \hline
\multirow{2}{*}{\textbf{Community}}            & 3 & 20, 45, 70 & 1.96E+05 & 1.66E+05 & 1.40E+05 & 2.45E+05 & 2.00E+05 & 1.65E+05 & 22.32\%  & 18.46\%  & 16.21\%  \\ \cline{2-12} 
                                               & 3 & 41, 66, 91 & 2.46E+05 & 2.13E+05 & 1.86E+05 & 3.00E+05 & 2.56E+05 & 2.16E+05 & 19.69\%  & 18.07\%  & 15.09\%  \\ \hline
\multirow{2}{*}{\textbf{Forest Hub and Spoke}} & 3 & 17, 42, 67 & 1.35E+05 & 9.58E+04 & 7.38E+04 & 3.36E+05 & 2.33E+05 & 1.49E+05 & 85.13\%  & 83.43\%  & 67.28\%  \\ \cline{2-12} 
                                               & 3 & 35, 60, 85 & 1.57E+05 & 1.16E+05 & 9.08E+04 & 3.72E+05 & 2.53E+05 & 1.68E+05 & 81.25\%  & 74.44\%  & 59.89\%  \\ \hline
\multirow{2}{*}{\textbf{Scale-free}}           & 3 & 17, 42, 67 & 1.89E+05 & 1.25E+05 & 9.89E+04 & 2.74E+05 & 1.98E+05 & 1.42E+05 & 36.55\%  & 45.35\%  & 35.80\%  \\ \cline{2-12} 
                                               & 3 & 33, 58, 83 & 2.13E+05 & 1.46E+05 & 1.20E+05 & 3.04E+05 & 2.24E+05 & 1.64E+05 & 35.55\%  & 42.13\%  & 31.42\%  \\ \hline
\multirow{2}{*}{\textbf{Hub \& Spoke}}         & 3 & 5, 30, 55  & 4.56E+05 & 7.19E+04 & 4.11E+04 & 7.88E+05 & 4.17E+05 & 1.66E+05 & 53.37\%  & 141.16\% & 120.87\% \\ \cline{2-12} 
                                               & 3 & 14, 39, 64 & 4.82E+05 & 8.07E+04 & 5.19E+04 & 9.01E+05 & 5.18E+05 & 2.35E+05 & 60.69\%  & 146.08\% & 127.51\% \\ \hline
\end{tabular}%
}
\end{table}

The difference, then, in the total number of messages is due to the agents' `light' communication incurred by their movement process. 
That is, each agent located on a node must obtain the `visited' status of the neighbouring nodes, via a ping signal, before determining its next move. As a node's `visited' status merely corresponds to a boolean flag, these extra messages do not really impact the amount data transfers. Moreover, once the maximum topology data is collected, the agents are dropped, and so are the messages they produce. Without the ping messages, the number of messages incurred in each round correspond to the number of mobile agents. While Trickle sends fewer messages, its messages are bigger hence consuming more bandwidth overall. 

\begin{figure}[H]
\centering
\includegraphics[scale=0.45]{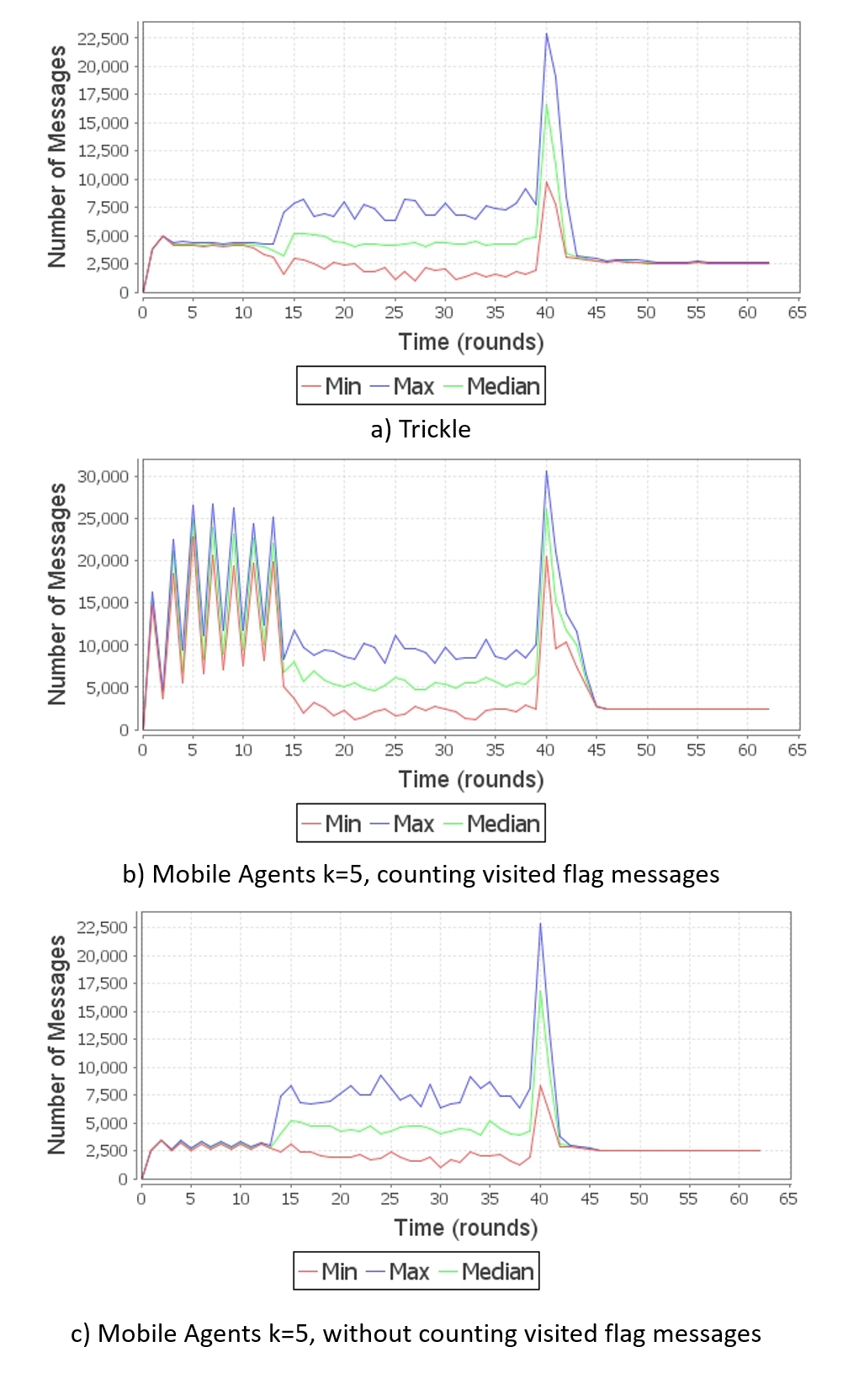}
\caption{Number of messages received by all nodes in the AS network, for $p_f=0.5$, scenario (13, 38, 63). The redundancy factor for Mobile Agents was $k=5$ and for Trickle $k=3$.}
\label{fig:nummessages}       
\end{figure}

Figure  \ref{fig:nummessages} shows the number of messages received by all nodes within the AS network, per each round. The tested scenario was (13, 38, 63), with nodes failing from round 13 to round 38, and then recovering until the end of the experiment. After the node recovery period, it is observed that received messages stabilise to a constant value of about 2500 after the self-healing process, for both Trickle and Mobile Agents. This is related with the number of links in each network (e.g. AS topology here) as they correspond to heartbeat messages for checking neighbour aliveness. This confirms that the network topology was recovered entirely. 

When including `ping' messages in the Mobile Agents' message count, it can be observed that the number of messages varies significantly with the agents' locations over time; more precisely with the degrees of visited nodes (as agents ping neighbouring nodes for their `visited' status, when deciding their next move). Such variations can be observed in Fig. \ref{fig:nummessages}-b, for Mobile Agents, during the data-collection period (before the `start' point, at round 13).  While agents are created during the node self-healing period, they are quickly discarded once visited nodes establish that they carry redundant information (Cf. subsec. \ref{subsubsubsec:mobile-agents-submodel} and \ref{subsubsec:self-healing}).    

\begin{figure}[H]
\centering
\includegraphics[width=\textwidth]{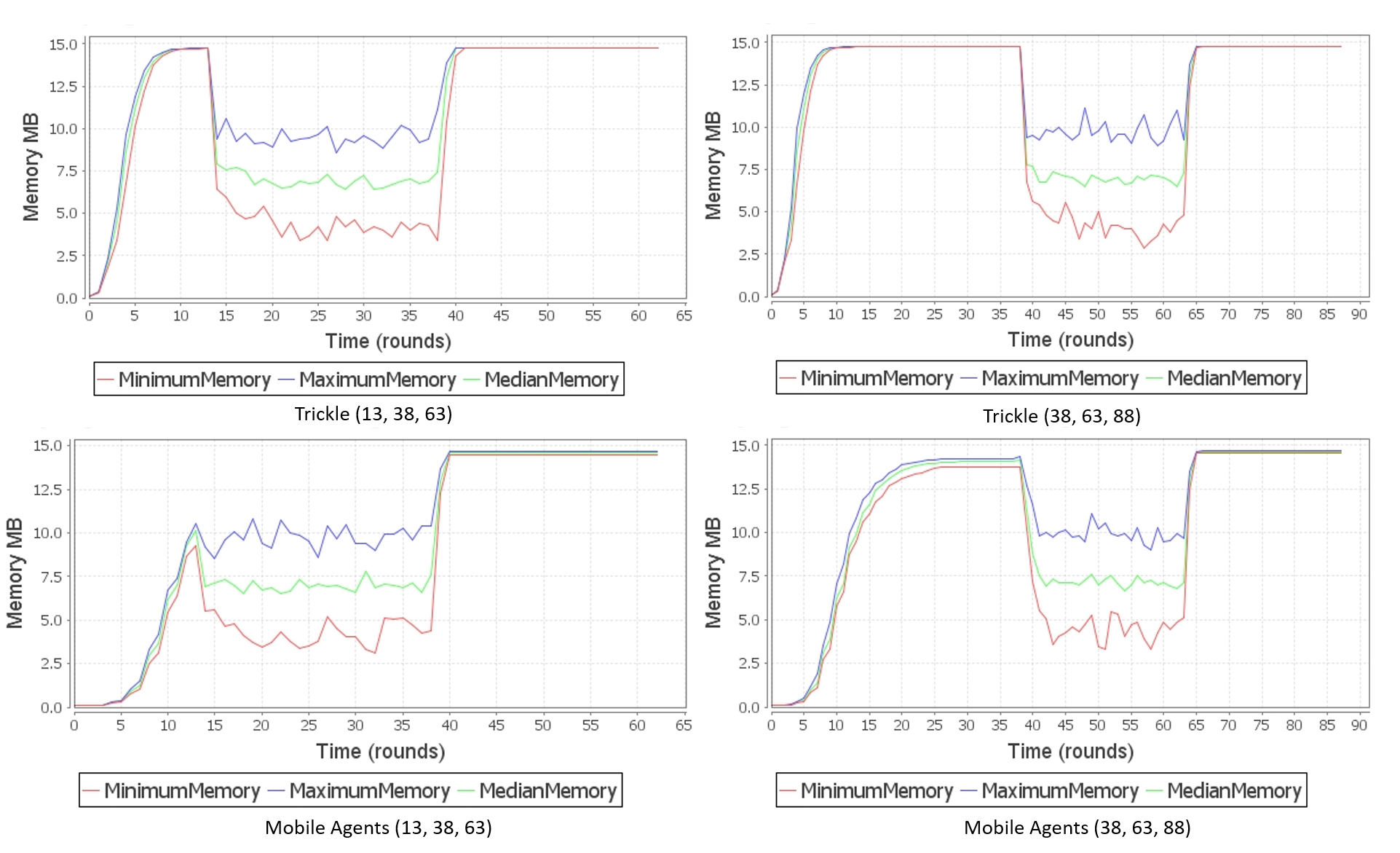}
\caption{Memory consumption for the AS network, with $p_f = 0.5$. Mobile Agents had redundancy factor $k = 5$, and Trickle $k = 3$. }
\label{fig:memory}       
\end{figure}

{Figure \ref{fig:memory} presents 
the increasing memory size for storing topology data in nodes, in Megabytes, for the AS network. In the upper part it is noted that nodes using Trickle collect the entire topology data about the network. When using Mobile Agents, even with $k=5$, nodes only acquire partial information about the network topology. A similar result is observed for the other  selected networks} {Table \ref{tbl:mem-cons-node} presents the integral over the entire experiment of the size of the topology data (in bytes) present in nodes (minimum, median and maximum).} Memory-wise, Mobile Agents consume less memory than Trickle, if we only consider the topology data stored in the nodes (Cf. Figure \ref{fig:memory} and Table \ref{tbl:mem-cons-node}). {This occurs because agents transport and deposit less data, and that onto a limited number of visited nodes, whereas Trickle imposes that all nodes send all their data to all of their neighbours.}

\begin{table}[H]
\centering
\caption{Memory consumption in all nodes.  The redundancy factor for Trickle was $k=3$}
\label{tbl:mem-cons-node}
\resizebox{\textwidth}{!}{%
\begin{tabular}{|l|c|c|c|c|c|c|c|c|c|c|c|}
\hline
\multicolumn{1}{|c|}{\multirow{2}{*}{\textbf{Network}}} &
  \multicolumn{1}{l|}{\multirow{2}{*}{\textbf{\begin{tabular}[c]{@{}l@{}}k for mobile\\ agents\end{tabular}}}} &
  \multicolumn{1}{l|}{\multirow{2}{*}{\textbf{\begin{tabular}[c]{@{}l@{}}Expr.\\ Conf.\end{tabular}}}} &
  \multicolumn{3}{c|}{\textbf{Trickle}} &
  \multicolumn{3}{c|}{\textbf{Mobile Agents}} &
  \multicolumn{3}{c|}{\textbf{Difference Percentage RPD}} \\ \cline{4-12} 
\multicolumn{1}{|c|}{} &
  \multicolumn{1}{l|}{} &
  \multicolumn{1}{l|}{} &
  \multicolumn{1}{l|}{\textbf{Max}} &
  \multicolumn{1}{l|}{\textbf{Median}} &
  \multicolumn{1}{l|}{\textbf{Min}} &
  \multicolumn{1}{l|}{\textbf{Max}} &
  \multicolumn{1}{l|}{\textbf{Median}} &
  \multicolumn{1}{l|}{\textbf{Min}} &
  \multicolumn{1}{l|}{\textbf{Max}} &
  \multicolumn{1}{l|}{\textbf{Median}} &
  \multicolumn{1}{l|}{\textbf{Min}} \\ \hline
\multirow{4}{*}{\textbf{AS19981231}}           & 3 & 13, 38, 63 & 3.65E+09 & 3.31E+09 & 2.94E+09 & 3.17E+09 & 2.78E+09 & 2.40E+09 & 14.05\% & 17.16\% & 20.48\% \\ \cline{2-12} 
                                               & 3 & 38, 63, 88 & 5.51E+09 & 5.14E+09 & 4.78E+09 & 4.95E+09 & 4.54E+09 & 4.15E+09 & 10.68\% & 12.36\% & 14.15\% \\ \cline{2-12} 
                                               & 5 & 13, 38, 63 & 3.65E+09 & 3.31E+09 & 2.94E+09 & 3.17E+09 & 2.79E+09 & 2.42E+09 & 14.13\% & 16.78\% & 19.38\% \\ \cline{2-12} 
                                               & 5 & 38, 63, 88 & 5.51E+09 & 5.14E+09 & 4.78E+09 & 4.99E+09 & 4.58E+09 & 4.19E+09 & 9.84\%  & 11.62\% & 13.14\% \\ \hline
\multirow{2}{*}{\textbf{Small-World}}          & 3 & 12, 37, 62 & 1.19E+08 & 1.10E+08 & 9.96E+07 & 1.08E+08 & 9.84E+07 & 8.70E+07 & 10.41\% & 10.96\% & 13.44\% \\ \cline{2-12} 
                                               & 3 & 25, 50, 75 & 1.51E+08 & 1.41E+08 & 1.31E+08 & 1.40E+08 & 1.30E+08 & 1.18E+08 & 7.83\%  & 8.80\%  & 9.77\%  \\ \hline
\multirow{2}{*}{\textbf{Community}}            & 3 & 20, 45, 70 & 1.36E+08 & 1.25E+08 & 1.12E+08 & 1.17E+08 & 1.04E+08 & 8.41E+07 & 14.64\% & 18.57\% & 28.13\% \\ \cline{2-12} 
                                               & 3 & 41, 66, 91 & 1.88E+08 & 1.76E+08 & 1.63E+08 & 1.65E+08 & 1.47E+08 & 1.19E+08 & 12.63\% & 17.94\% & 31.17\% \\ \hline
\multirow{2}{*}{\textbf{Forest Hub and Spoke}} & 3 & 17, 42, 67 & 1.07E+08 & 9.96E+07 & 8.62E+07 & 9.49E+07 & 8.49E+07 & 6.91E+07 & 12.02\% & 15.94\% & 22.01\% \\ \cline{2-12} 
                                               & 3 & 35, 60, 85 & 1.39E+08 & 1.32E+08 & 1.21E+08 & 1.25E+08 & 1.08E+08 & 8.52E+07 & 11.07\% & 20.37\% & 34.31\% \\ \hline
\multirow{2}{*}{\textbf{Scale-free}}           & 3 & 17, 42, 67 & 1.17E+08 & 1.08E+08 & 9.90E+07 & 1.03E+08 & 9.21E+07 & 7.32E+07 & 13.17\% & 16.29\% & 29.98\% \\ \cline{2-12} 
                                               & 3 & 33, 58, 83 & 1.48E+08 & 1.40E+08 & 1.30E+08 & 1.32E+08 & 1.18E+08 & 9.47E+07 & 11.49\% & 17.43\% & 31.56\% \\ \hline
\multirow{2}{*}{\textbf{Hub \& Spoke}}         & 3 & 5, 30, 55  & 9.41E+07 & 8.88E+07 & 6.47E+07 & 8.86E+07 & 8.40E+07 & 5.92E+07 & 5.97\%  & 5.48\%  & 8.88\%  \\ \cline{2-12} 
                                               & 3 & 14, 39, 64 & 1.11E+08 & 1.05E+08 & 8.56E+07 & 1.04E+08 & 9.87E+07 & 7.47E+07 & 6.45\%  & 6.30\%  & 13.71\% \\ \hline
\end{tabular}%
}
\end{table}

If we consider the total memory consumption, including both the topology data stored in nodes and in mobile agents, then Trickle consumes less memory overall. {Table \ref{tbl:totalmemcons} presents the total memory consumed by the two data collection algorithms. Trickle is impacted only by the storage of node topology. 
Mobile Agents involve memory consumption on the nodes both for storing topology data and the mobile agents themselves. That said, memory overheads induced by the mobile agents are temporary and limited to the data-collection and dissemination periods.}

\begin{table}[H]
\centering
\caption{Total memory consumed by Trickle and Mobile Agents. The redundancy factor for Trickle was $k=3$}
\label{tbl:totalmemcons}
\resizebox{\textwidth}{!}{%
\begin{tabular}{|l|c|c|c|c|c|c|c|c|c|c|c|}
\hline
\multicolumn{1}{|c|}{\multirow{2}{*}{\textbf{Network}}} &
  \multirow{2}{*}{\textbf{\begin{tabular}[c]{@{}c@{}}k for mobile\\ agents\end{tabular}}} &
  \multirow{2}{*}{\textbf{\begin{tabular}[c]{@{}c@{}}Exp.\\ Conf.\end{tabular}}} &
  \multicolumn{3}{c|}{\textbf{Trickle}} &
  \multicolumn{3}{c|}{\textbf{Mobile Agents}} &
  \multicolumn{3}{c|}{\textbf{Difference Percentage RPD}} \\ \cline{4-12} 
\multicolumn{1}{|c|}{} &
   &
   &
  \textbf{Max} &
  \textbf{Median} &
  \textbf{Min} &
  \textbf{Max} &
  \textbf{Median} &
  \textbf{Min} &
  \textbf{Max} &
  \textbf{Med} &
  \textbf{Min} \\ \hline
\multirow{4}{*}{\textbf{AS19981231}} &
  3 &
  13, 38, 63 &
  3.65E+09 &
  3.31E+09 &
  2.94E+09 &
  5.00E+09 &
  4.13E+09 &
  3.32E+09 &
  31.27\% &
  22.16\% &
  11.98\% \\ \cline{2-12} 
 &
  3 &
  38, 63, 88 &
  5.51E+09 &
  5.14E+09 &
  4.78E+09 &
  7.63E+09 &
  6.47E+09 &
  5.46E+09 &
  32.30\% &
  22.78\% &
  13.31\% \\ \cline{2-12} 
 &
  5 &
  13, 38, 63 &
  3.65E+09 &
  3.31E+09 &
  2.94E+09 &
  5.15E+09 &
  4.30E+09 &
  3.49E+09 &
  34.25\% &
  26.14\% &
  16.90\% \\ \cline{2-12} 
 &
  5 &
  38, 63, 88 &
  5.51E+09 &
  5.14E+09 &
  4.78E+09 &
  8.08E+09 &
  6.83E+09 &
  5.76E+09 &
  37.91\% &
  28.12\% &
  18.65\% \\ \hline
\multirow{2}{*}{\textbf{Small-World}} &
  3 &
  12, 37, 62 &
  1.19E+08 &
  1.10E+08 &
  9.96E+07 &
  1.69E+08 &
  1.46E+08 &
  1.21E+08 &
  34.44\% &
  28.17\% &
  19.42\% \\ \cline{2-12} 
 &
  3 &
  25, 50, 75 &
  1.51E+08 &
  1.41E+08 &
  1.31E+08 &
  2.21E+08 &
  1.93E+08 &
  1.68E+08 &
  37.70\% &
  31.03\% &
  25.24\% \\ \hline
\multirow{2}{*}{\textbf{Community}} &
  3 &
  20, 45, 70 &
  1.36E+08 &
  1.25E+08 &
  1.12E+08 &
  1.81E+08 &
  1.43E+08 &
  1.09E+08 &
  28.34\% &
  13.46\% &
  2.51\% \\ \cline{2-12} 
 &
  3 &
  41, 66, 91 &
  1.88E+08 &
  1.76E+08 &
  1.63E+08 &
  2.44E+08 &
  2.04E+08 &
  1.56E+08 &
  26.32\% &
  14.63\% &
  4.55\% \\ \hline
\multirow{2}{*}{\textbf{Forest Hub \& Spoke}} &
  3 &
  17, 42, 67 &
  1.07E+08 &
  9.96E+07 &
  8.62E+07 &
  1.47E+08 &
  1.18E+08 &
  8.57E+07 &
  31.61\% &
  16.55\% &
  0.65\% \\ \cline{2-12} 
 &
  3 &
  35, 60, 85 &
  1.39E+08 &
  1.32E+08 &
  1.21E+08 &
  1.80E+08 &
  1.41E+08 &
  1.03E+08 &
  25.45\% &
  6.86\% &
  15.80\% \\ \hline
\multirow{2}{*}{\textbf{Scale-free}} &
  3 &
  17, 42, 67 &
  1.17E+08 &
  1.08E+08 &
  9.90E+07 &
  1.57E+08 &
  1.29E+08 &
  9.59E+07 &
  28.67\% &
  17.63\% &
  3.11\% \\ \cline{2-12} 
 &
  3 &
  33, 58, 83 &
  1.48E+08 &
  1.40E+08 &
  1.30E+08 &
  1.90E+08 &
  1.57E+08 &
  1.19E+08 &
  25.14\% &
  11.58\% &
  9.12\% \\ \hline
\multirow{2}{*}{\textbf{Hub \& Spoke}} &
  3 &
  5, 30, 55 &
  9.41E+07 &
  8.88E+07 &
  6.47E+07 &
  1.37E+08 &
  1.16E+08 &
  6.41E+07 &
  37.30\% &
  26.45\% &
  0.86\% \\ \cline{2-12} 
 &
  3 &
  14, 39, 64 &
  1.11E+08 &
  1.05E+08 &
  8.56E+07 &
  1.61E+08 &
  1.36E+08 &
  8.69E+07 &
  37.03\% &
  25.64\% &
  1.41\% \\ \hline
\end{tabular}%
}
\end{table}

\subsection{Discussion}

Experimental results showed that the proposed self-healing approach, with both data distribution variants (Trickle and Mobile Agents), was able to recover the network, for all topologies, up to a maximum probability of node failure ($p_f = 0.25$ or $0.5$, depending on the network topology). Results also showed how topology impacts recovery success rates. Namely, topologies containing nodes with low degrees (e.g. one or two links per node, in the Scale-free and Hub \& Spoke examples) were more prone to losing nodes than topologies where all nodes were well-connected. This can be expected as marginal nodes only have one or two neighbours that can recover them. When these neighbours also fail, at high $p_f$ rates, the marginal nodes are lost (e.g. Fig. \ref{fig:missingnodes}).  

Marginal nodes are also less likely to be visited and hence known by their neighbours, further increasing their chances of being lost. {In the AS network}, increasing the redundancy factor for Mobile Agents  (i.e. from $k=3$ to $k=5$) ensured that local information was disseminated to all neighbourhoods, which could then be recreated when nodes failed. This engendered more agent traffic yet still outperformed Trickle in terms of lesser bandwidth overheads.

At the same time, the two tested {data collection approaches} were shown to achieve different degrees of topological information dissemination, and to incur different kinds of resource overheads, in terms of memory consumption, communication bandwidth and message numbers. {Trickle} was able to disseminate the complete topology information rather quickly (e.g. less than 15 rounds for the largest selected network -- `AS', with 512 nodes), and hence to ensure complete network recovery. Mobile Agents were shown to disseminate slightly less information, at a lower pace (e.g. less than 40 rounds for the `AS' network), yet this proves \textit{not} to be essential for the network recovery process. Hence, Mobile Agents achieved equivalent network self-healing results, while incurring lower bandwidth overheads (e.g. up to 100\% difference for the `AS' network). This was because agents carry less topological information, hence incurring smaller messages. Still, as they manage to disseminate local information within network neighbourhoods, this suffices to recover local node failures. Similarly, while Mobile Agents take more time than Trickle to collect maximum topology information (e.g. 38 vs 13 rounds for `AS' network), they could recover the entire network without collecting the maximum topology data (e.g. even when nodes started failing at round 13 in the `AS' case).

At the same time, agents consume extra memory when stored by nodes (e.g. up to 38\% more for the `AS' network) and incur more messages as they ping neighbours before moving (e.g. up to 100\% more for the `AS' network, yet still ensuing less bandwidth consumption overall).  
As agents incur extra memory and bandwidth consumption, we limited their population and their usage to only those periods during which data-collection and dissemination was needed. Hence, nodes discard agents when they start carrying redundant information, as they are not needed for the self-healing process. Future work will explore cases where agents are recreated in reaction to intended topological changes that need to be disseminated. With this approach, Mobile Agents incur more memory usage than Trickle during data-collection time, but lesser once topology information is available in the nodes and the agents are discarded (e.g. up to 20\% less for the `AS' network, once agents are dropped). 

Overall, these results show, on the one hand, the ability of the proposed self-healing method to recover a network's topology from node failures; and, on the other hand, that collecting and disseminating only partial topological information to network neighbourhoods (e.g. as with the proposed Mobile Agents approach) suffices to maintain the topology, while incurring lesser bandwidth overheads than exhaustive broadcasting methods (e.g. Trickle).  Future work will further explore the extent of local information that is necessary to recover a network from failures that occur with various frequencies; as well as the best methods for collecting and disseminating such local information.

\section{Conclusions and Future Work}
\label{sec:conclusions}

This paper presented a self-healing approach for recovering a network's topology from node failures. This differs from existing solutions in that it recreates missing nodes and their connections based on topology information, rather than merely reconnecting remaining nodes. The approach relies on two decentralised mechanisms: i) data collection and dissemination across network nodes, on the network topology to be maintained; and, ii) node recovery and re-connection process, by one of the neighbouring nodes, which also prompts all neighbours to share topological information with node replicas.

Topology information was obtained via two data-collection and dissemination approaches. Firstly, we proposed a Mobile Agents-based algorithm (adapted from previous work); and secondly, we adopted a reference gossiping protocol from the literature (Trickle). 
In all cases, we only considered experimental scenarios where node failure rates were lower than the time taken by the network self-healing processes.

Overall, both variants were able to recover the network topology, up to a maximum failure probability rate, which depended on the topology maintained. This validates the proposed network recovery method and highlights its generality with respect to the data collection and dissemination method.  

Moreover, experimental results allowed us to compare the proposed Mobile Agents approach (collecting partial topological data) with the reference Trickle approach (broadcasting the entire topology) -- in terms of memory consumption, communication bandwidth and number of exchanged messages. While mobile Agents are significantly lighter in terms of bandwidth consumption, they consume more memory and initiate more messages than Trickle, during the data-collection process. These insights allow system designers to select the most suitable variant depending on their particular application context. 


%
Future work will extend the proposed data-collection algorithms so as to minimise the amount of data exchanged between nodes, by pruning transmitted topology information to a specific number of hops. The selected neighbourhood size should suffice to recover the network when nodes fail with a specific frequency. 
Additionally, future work will evaluate the proposal's scaling behaviour, when targeting larger networks (i.e. various orders of magnitude). 

Finally, we intend to apply the proposed approach to container networks. A container network is an overlay network \cite{Makowski2019}, where links between two nodes are virtual, or logical, over multiple links of a physical network (e.g. a peer-to-peer network over the internet) \cite{huang2017mobile}. A container represents a node that provides storage or other services, and it works like an application over the operating system. This offers portability, flexibility and short startup times. Container networks applications include the creation of digital market places \cite{Makowski2019} or data-streaming applications \cite{marinescu2017cloud}. 
This is a suitable application for our proposal as it can manage topological changes occurring due to container failures or node disconnections via denial-of-service attacks \cite{feldmann2020survey}. 


\bibliographystyle{spmpsci}      
\bibliography{Manuscript}   

%
%
 
\end{document}